\documentclass[preprint,aps,prd,showpacs,superscriptaddress,nofootinbib]{revtex4}
\usepackage{mathbbold}
\usepackage{amsfonts}
\usepackage{mathrsfs}
\usepackage{graphicx}
\usepackage{amsmath}
\usepackage{amssymb}
\usepackage{bm}
\usepackage{bbm}

\def\OMIT#1{}

\newcommand{\nn}{\nonumber}

\newcommand{\beq}{\begin{equation}}
\newcommand{\eeq}{\end{equation}}
\newcommand{\bqa}{\begin{eqnarray}}
\newcommand{\eqa}{\end{eqnarray}}

\begin{document}


\title{\mbox{}\\[10pt]
Bridging light-cone and NRQCD approaches: asymptotic behavior of ${\bm B}_{\bm c}$
electromagnetic form factor}

\author{Yu Jia\footnote{E-mail: jiay@ihep.ac.cn}}
\affiliation{Institute of High Energy Physics, Chinese Academy of
Sciences, Beijing 100049, China\vspace{0.2cm}}
\affiliation{Theoretical Physics Center for Science Facilities,
Chinese Academy of Sciences, Beijing 100049, China\vspace{0.2cm}}

\author{Jian-Xiong Wang\footnote{E-mail: jxwang@ihep.ac.cn}}
\affiliation{Institute of High Energy Physics, Chinese Academy of
Sciences, Beijing 100049, China\vspace{0.2cm}}
\affiliation{Theoretical Physics Center for Science Facilities,
Chinese Academy of Sciences, Beijing 100049, China\vspace{0.2cm}}

\author{Deshan Yang\footnote{E-mail: yangds@gucas.ac.cn}}
\affiliation{College of Physical Sciences, Graduate University of
Chinese Academy of Sciences, Beijing 100049, China\vspace{0.2cm}}

\date{\today}
\begin{abstract}
This work aims at illustrating that, for a class of leading-twist
hard exclusive reactions involving two heavy quarkonia, the light-cone approach,
when equipped with the strategy of refactorization of the light-cone distribution
amplitude of quarkonium,
can be employed to elegantly reproduce the corresponding predictions
made in the nonrelativistic QCD (NRQCD) factorization approach, order by order in
perturbative expansion.
Taking the electromagnetic form factor of the $B_c$ meson at
large momentum transfer, $Q^2$, as a concrete example,
we compare the results obtained from both NRQCD-based and light-cone-based calculations
through the next-to-leading order (NLO) in $\alpha_s$,
while at the leading order (LO) in both velocity and $1/Q^2$ expansion, and explicitly
confirm their mutual agreement. As a byproduct, we apply our NLO result to
explore certain features about the asymptotic behavior of the heavy-light meson form factor.
We also address the major theoretical obstacles that prevent us
from establishing an analogous equivalence between these two approaches
for the double charmonium production process of phenomenological interest,
$e^+e^-\to J/\psi+\eta_c$.

\end{abstract}

\pacs{\it 12.38.-t, 12.38.Bx, 12.39.Hg, 13.40.Gp, 14.40.Pq}
\maketitle


\section{Introduction}

The utility of perturbative QCD crucially rests upon the idea of
factorization. For a typical QCD process with large momentum
transfer (either inclusive or exclusive), factorization provides an
essential tool to systematically separate the short-distance,
perturbatively calculable effects from the long-distance, yet
universal, nonperturbative effects. For a hard exclusive processes
involving a few hadrons, the well-known collinear factorization
(also referred to as {\it light-cone approach} in
literature)~\cite{Lepage:1980fj,Chernyak:1983ej}, fulfills such a
separation by expressing the amplitude as the convolution of the
hard-scattering part with the nonperturbative yet universal
light-cone distribution amplitudes (LCDAs) of the corresponding
hadrons. The classic applications of light-cone factorization are
exemplified by the $\pi\!-\!\gamma$ transition form factor and $\pi$
electromagnetic (EM) form factor~\cite{Lepage:1980fj,Chernyak:1983ej},
and the nonleptonic $B$ meson decays~\cite{Beneke:1999br,Beneke:2000ry},
to which a vast amount of literature has been devoted.

Recent advancement of the high-luminosity high-energy collider facilities
renders the study of hard exclusive processes involving heavy quarkonium
a fertile research frontier.
Perhaps a great amount of interests toward this topic
have been triggered by the observation of
several double-charmonium production processes in two $B$ factories
several years ago~\cite{Abe:2002rb,Aubert:2005tj}.

For a hard exclusive process comprising entirely of light hadrons,
such as pion EM form factor, only the hard momentum transfer $Q$ and the
nonperturbative QCD scale, $\Lambda_{\rm QCD}$,  play dynamical
roles. By contrast, hard exclusive reactions involving quarkonium
bring forth much more complexity as a number of new dynamic scales
have come into play altogether. Among the important scales inherent
to a heavy quarkonium, are the heavy quark mass, $m$, the typical
momentum of heavy quark, $mv$, and the typical binding energy of a
quarkonium, $mv^2$, where $v$ denotes the typical velocity of heavy
quark inside a quarkonium. As a consequence, it is theoretically
more challenging to analyze the hard reactions involving heavy
quarkonia than those involving only light hadrons.

Provided that the characteristic momentum transfer, $Q$, is much greater than $m$,
the light-cone approach undoubtedly remains to be a valid and powerful method,
since it is a formalism based on the expansion in powers of $1/Q$.
Nevertheless, there also exists another influential factorization approach,
the {\it NRQCD factorization}~\cite{Bodwin:1994jh},
which combines the effective-field-theory (EFT) machinery and the hard-scattering factorization.
Tailor-made to tackle quarkonium production and decay processes,
this method explicitly exploits the nonrelativistic nature of quarkonium
by factoring the perturbative quantum fluctuations of distance $1/m$ or shorter,
from the nonperturbative effects governing the transition of a heavy quark pair
into a physical quarkonium, which occurs at a distance of $1/mv$ or longer.
As a consequence, NRQCD factorization allows one to express the amplitude of an exclusive
quarkonium production reaction
in terms of an infinite sum of products of
short-distance coefficients and the vacuum-to-quarkonium
NRQCD matrix elements, whose importance has been organized
by the powers of $v$.

Both factorization frameworks, despite spotlighting different dynamic aspects,
are well-established approaches derived from the first principles of QCD.
They are commonly viewed as two drastically different, and,
even repulsive methods.
For example, in recent years, intensive investigations
on the reaction $e^+e^-\to J/\psi+\eta_c$ have been conducted from both
frameworks~\cite{Braaten:2002fi,Liu:2002wq,Hagiwara:2003cw,Zhang:2005cha,Gong:2007db,Ma:2004qf,Bondar:2004sv,Braguta:2008tg},
and there have been some disputes on which approach
is superior when applied to the hard exclusive reaction involving quarkonium~\cite{Bondar:2004sv,Bodwin:2006dm}.

However, it seems fair to state that, when coping with hard exclusive process
involving quarkonium, each approach has its own strength and weakness.
For example, in NRQCD factorization, the short-distance coefficients
at each order in $\alpha_s$ would inevitably contain large logarithms of
the ratio of two vastly different scales, $Q$
({\it e.g.}, the center-of-mass energy for quarkonium$+\gamma$ or
double-charmonium production in $e^+e^-$ annihilation) and $m$.
To improve the stability of perturbative expansion,
a systematic disentanglement of these two scales is clearly called for.
On the other hand, the light-cone approach, being a
general formalism applicable to any type of hadrons,
usually does not sufficiently exploit the nonrelativistic
trait of quarkonium. That is, the LCDAs of quarkonium are often determined
on purely phenomenological ground~\cite{Bondar:2004sv,Bodwin:2006dm,Braguta:2006wr,Hwang:2008qi}.
The unsatisfactory feature of this approach is that, quarkonium LCDAs inevitably
encompass the collinear degrees of freedom of vastly varying virtualities,
some of which may be high enough to be perturbatively calculable and
separable from the remaining parts.

Recently it has been realized that these two methods need not be
regarded solely as rivals, rather they can be tied coherently to
compensate the shortcoming of each other. In Ref.~\cite{Jia:2008ep},
it is suggested that the light-cone approach can be utilized to
refactorize the NRQCD short-distance coefficients for a class of
single-quarkonium production processes, exemplified by
$\eta_b\!-\!\gamma$ form factor and Higgs bosn decay to
$\Upsilon+\gamma$, by which the scales $Q$ and $m$ can get fully
disentangled. Consequently, the collinear logarithms of type
$\alpha_s^n\ln^n (Q^2/m^2)$ can be readily identified and resummed in
this framework~\footnote{It is interesting to remark that,
resummation of the large kinematic logarithms for the
above-mentioned exclusive single-quarkonium production processes
have also been performed by Shifman and Vysotsky thirty years
ago~\cite{Shifman:1980dk}, long before the inception of the NRQCD factorization
approach.}. Furthermore, it is envisaged in \cite{Jia:2008ep} that
the light-cone approach can serve as an alternative means to
reproduce the NRQCD factorization predictions at the leading order
in $1/Q$ and order by order in $\alpha_s$. Therefore this
refactorization procedure can spare a great amount of labors
compared with the direct higher-order computations in NRQCD
framework.

There has already existed another interesting development from a
different but complimentary perspective, which explores the
possibility of further factoring the LCDAs of
quarkonium~\cite{Ma:2006hc}. The underlying idea is that, since the
virtualities of the collinear modes encoded in quarkonium LCDA can
range from ${\mathcal O}(m^2)$ down to ${\mathcal O}(m^2 v^2)$ or lower, it
seems profitable to identify and separate the shorter-distance
collinear quantum fluctuations out of the quarkonium LCDA. Thus, it is suggested in
\cite{Ma:2006hc} that, the LCDAs of a quarkonium can be matched onto
an infinite sum of the product of the perturbatively-calculable
universal jet functions and the vacuum-to-quarkonium NRQCD matrix
elements, organized by $v$ expansion. At the lowest-order (LO) in
$v$, the corresponding jet functions associated with the twist-2
LCDAs of the S-wave quarkonia, such as $\eta_c$, $J/\psi$ and $B_c$,
have been calculated through the next-to-leading-order (NLO) in
$\alpha_s$~\cite{Ma:2006hc,Bell:2008er}.

Both versions of refactorization, operating either on NRQCD
short-distance coefficients~\cite{Jia:2008ep},
or on quarkonium LCDAs~\cite{Ma:2006hc,Bell:2008er}, aim to achieve an optimized scale separation
through bridging the light-cone and NRQCD approaches together.
Though motivated from different considerations
in curing the particular shortcoming of the respective factorization approach,
both of them in fact lead to equivalent final predictions,
that is, the exclusive reaction amplitude will be expressible in the form of
a hard scattering kernel convoluted with the pertubative jet function, then
multiplied by the nonperturbative vacuum-to-quarkonium NRQCD matrix element.
Practically speaking, the second version of refactorization strategy~\cite{Ma:2006hc},
which highlights the universality of the perturbatively-calculable jet functions,
is presumably more convenient to employ, and we will follow this perspective in this work.

The main motif of this work is to verify the {\it correctness} and
{\it effectiveness} of the refactorization strategy for a class of
{\it leading-twist} hard exclusive processes involving {\it two}
quarkonia, which is more nontrivial and interesting than those
single-quarkonium production processes considered in
\cite{Jia:2008ep}. Specifically,
by employing the NRQCD approach and the light-cone approach separately,
we will investigate the EM form
factor of the $B_c$ meson at asymptotically large momentum transfer.
In spite of lacking urgent phenomenological incentive,
this process can serve as an ideal theoretical laboratory to
corroborate our understanding in a nontrivial way. We will
explicitly verify that, to the accuracy at LO in $v$ but through NLO
in $\alpha_s$, the light-cone approach, when equipped with the
machinery of refactorization, can be utilized in an effective and
elegant manner to reproduce the asymptotic NRQCD predictions for the
$B_c$ EM form factor. Aside from the calculational advantage, the
refactorization strategy may also shed light on how to ameliorate
the severe scale dependence associated with the NLO NRQCD
predictions observed in various quarkonium production processes.

Apart from its great utility, however, it is worth emphasizing
one important caveat for this refactorization program.
Due to some long-standing problems in the collinear factorization framework,
the refactorization strategy, at its present formulation, can not be applied successfully
to any types of hard exclusive reactions involving quarkonium.
As we will see later, the phenomenologically interesting double-charmonium
production process, $e^+e^-\to J/\psi+\eta_c$, constitutes such a very example.

The rest of the paper is structured as follows.
In Sec.~\ref{Def:Bc:FF} we define the electromagnetic
form factor of the $B_c$ meson.
In Sec.~\ref{Bc:FF:NRQCD:fac}, we compute the
$B_c$ EM form factor through the NLO in $\alpha_s$ while at the LO in velocity
within NRQCD factorization framework.
In Sec.~\ref{Bc:FF:light:cone:fac},
we reinvestigate the $B_c$ form factor at large $Q^2$
in the framework of collinear factorization that
implements the refactorization strategy, to the
leading power in $1/Q^2$ and $v^2$, yet through the NLO in $\alpha_s$.
We then compare the light-cone-based prediction and the asymptotic NRQCD result
through the NLO in $\alpha_s$,
and explicitly establish their equivalence at large $Q^2$.
In Sec.~\ref{heavy-light:meson:EM:FF},
we apply our knowledge of the NLO expression for the $B_c$ form factor
to discuss certain features of the EM form factor of a heavy-light meson like $B^+$,
and speculate on possible refactorization procedure suitable to this situation.
In Sec.~\ref{chall:Jpsi+etac:EM:FF},
we discuss the peculiar asymptotic behavior of the NLO NRQCD prediction to
the reaction $e^+e^-\to J/\psi+\eta_c$, and briefly address the
inapplicability of the refactorization strategy to this case.
Finally we summarize in Sec.~\ref{summary}.

\section{Definition of the $\bm{B}_{\bm c}$ electromagnetic form factor}
\label{Def:Bc:FF}

\begin{figure}
\begin{center}
\includegraphics[height=4 cm]{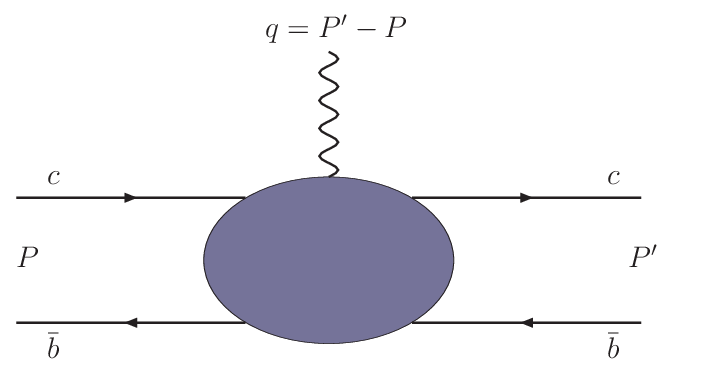}
\caption{
Schematic depiction of the quark transition process
relevant to the $B_c$ EM form factor.
}
\label{Fig:sketch:Bc:EM:form:factor}
\end{center}
\end{figure}

Imagine a $B_c$ meson struck by an electromagnetic probe undergoes an
elastic scattering.
The information of this elastic scattering is encoded in the
EM form factor of $B_c$, dubbed $F(Q^2)$ in this work.
It can be introduced in full analogy with the EM form factor of the charged
pion:
\bqa
\langle B_c^+(P^\prime)\vert J^\mu_{\rm em}\vert B_c^+(P)\rangle= F(Q^2)
(P+P^\prime)^\mu\,,
\label{eq:emff}
\eqa
where the electromagnetic current $J^\mu_{\rm em}\equiv \sum e_q
\bar q\gamma^\mu q$ ($e_q=+{2\over 3}$ for up-type quarks and
$e_q=-{1\over 3}$ for down-type quarks), and the momentum carried by
the EM probe is $q=P^\prime-P$, with $Q^2 \equiv -q^2>0$ signifying
the momentum transfer.
The structure of (\ref{eq:emff}) is uniquely determined by
Lorentz invariance and the EM current conservation.
A cartoon that depicts the $B_c$ form
factor is sketched in Fig.~\ref{Fig:sketch:Bc:EM:form:factor}.

Throughout this work, we will be primarily interested in the asymptotic behavior of the form factor
in the limit of $Q^2\gg M^2_{B_c}$~\footnote{At present there are no experimental facilities
which can directly measure this form factor at large spacelike momentum transfer. The next
generation $e^+e^-$ machines, {\it e.g.} the prospective International Linear Collider,
or the super-high-luminosity machine operating at the $Z^0$ pole,
may be possible to access this form factor in the time-like region.}. In such a situation,
both NRQCD factorization and collinear factorization are valid theoretical tools
to predict this form factor.
Although the dimensionless form factor itself is a Lorentz scalar,
the physical picture is probably more transparent if the Breit frame is specifically
borne in mind.
In such a frame, a $B_c$ meson moving
very fast along the $\hat{z}$ axis gets hit by a highly virtual photon,
then heads back along the $-\hat{z}$ axis with equal speed.

\section{EM form factor of $\bm{B}_{\bm c}$  in NRQCD factorization}
\label{Bc:FF:NRQCD:fac}

In a hard exclusive process involving quarkonium, a pair of heavy
quark and an antiquark have to be created/annihilated in short
distance; and in order to have a significant probability to
form/disintegrate a quarkonium, the typical relative motion between
the quark and antiquark should be necessarily slow. The first condition
guarantees that the asymptotic freedom can be invoked to compute the
hard-scattering quark amplitude in perturbation theory. The second
condition implies that, this hard-scattering amplitude is
insensitive to small change of the relative momentum of the
pair, thus one may expand the amplitude in power series of $v$,
subsequently absorb those $v$-dependent factors and
the quarkonium wave function into the nonperturbative matrix
elements, which are dictated by the quarkonium binding mechanism.
NRQCD factorization then naturally arises from this physical picture.

At the LO in $v$, the $B_c$ EM form factor in NRQCD factorization
can be written in the following form:
\bqa
F_{\rm NRQCD}(Q^2)&=& C(Q;m_c,m_b) \left( { f_{B_c}^{(0)} \over \sqrt{2 N_c}} \right)^2 + {\cal
O}(v^2)\,,
\label{LO:NRQCD:factorization:formula}
\eqa
where $C(Q;m_c,m_b)$ denotes the corresponding short-distance coefficient.
$f_{B_c}^{(0)}$ represents the nonperturbative vacuum-to-$B_c$ NRQCD matrix element,
which is defined through
\bqa
f_{B_c}^{(0)} &\equiv &  \sqrt{2\over M_{B_c}} \left|\langle 0 |\chi_b^\dagger \psi_c| B^+_c \rangle \right|
\equiv   \sqrt{2\over M_{B_c}} \left|\langle B^+_c | \psi_c^\dagger \chi_b| 0 \rangle \right|
= \sqrt{4N_c\over M_{B_c}}\,\psi_{B_c}(0).
\label{decay:constant:LO:alphas}
\eqa
Here $N_c=3$ is the number of the colors, $\psi_c$ and $\chi_b$
denote the Pauli spinor fields associated with the $c$ and $b$
flavors, respectively. The $B_c$ state in the above NRQCD matrix
elements is normalized non-relativistically. As an alert reader may
readily tell, $f_{B_c}^{(0)}$ in fact is nothing but the $B_c$ decay
constant rephrased in the NRQCD context~\footnote{Note that our
convention of defining the decay constant of a charged pseudoscalar
meson is such that $f_\pi=132 $ MeV.}. The superscript $0$ reminds
that this identification is accurate only at the lowest order in
$\alpha_s$ and $v$. As displayed in
(\ref{decay:constant:LO:alphas}), this entity can also be linked
with $\psi_{B_c}(0)$, the familiar Schr\"{o}dinger wave function at
the origin for $B_c$. Furthermore, we will assume $M_{B_c}=m_c+m_b$
everywhere in this work, which is legitimate
to the LO accuracy in $v$ expansion.

As stressed before, we will concentrate on the kinematic situation
$Q\gg m_{c,b}\gg \Lambda_{\rm QCD}$. The NRQCD short-distance coefficient,
$C(Q;m_c,m_b)$, can then be deduced reliably through the
perturbative matching procedure. Consequently, it
can be expanded in power series of $\alpha_s$ as $C=C^{(0)}+
{\alpha_s \over \pi} C^{(1)}+\cdots$. Accordingly, the NRQCD
prediction to the $B_c$ form factor can be organized as
\bqa
F_{\rm NRQCD} = F_{\rm NRQCD}^{(0)}+{\alpha_s(\mu_R^2)\over \pi} F_{\rm NRQCD}^{(1)}+\cdots.
\label{NRQCD:sep:LO:NLO}
\eqa

\subsection{NRQCD prediction at LO in $\alpha_s$}

At tree level, there are only four Feynman diagrams for the
quark-transition process $\gamma^*+c\bar{b}({}^1S_0^{(1)},P)\to
c\bar{b}({}^1S_0^{(1)},P^\prime)$. They can be obtained by
replacing the shaded ellipse in
Fig.~\ref{Fig:sketch:Bc:EM:form:factor} with a single gluonic
exchange between $c$ and $\bar{b}$ quarks, as well as by
intersecting the EM current either to the $c$ or $b$ quark line.
Assuming that both constituents in each pair have vanishing relative
motion, it then is a straightforward exercise to deduce $C^{(0)}$:
\bqa
C^{(0)}(Q;m_c,m_b) &=& 4\pi C_F \alpha_s(\mu_R^2)
\bigg(e_c {\bar{x}_0 Q^2+ 2x_0 M^2_{B_c}\over
Q^4 \bar{x}_0^3}-
e_b {x_0 Q^2+ 2 \bar{x}_0 M^2_{B_c}\over Q^4 x_0^3}\bigg).
\label{NRQCD:short:coeff:LO:physical:Bc}
\eqa
For abbreviation, we have defined $x_0\equiv {m_c\over M_{B_c}}$,
and ${\bar x}_0 \equiv 1-x_0={m_b\over M_{B_c}}$~\footnote{
Throughout this paper we adopt the notation that $\bar{x} \equiv 1-x$ for any momentum fraction $x$.}.
$C_F={N_c^2-1\over 2 N_c}$ denotes the Casmir of the
fundamental representation for $SU(N_c)$ group.

In order to make contact with the leading-twist light-cone
predictions, which will be reported in next section, we are
particularly interested in the asymptotic behavior of
Eq.~(\ref{NRQCD:short:coeff:LO:physical:Bc}) in the limit $Q \gg
M_{B_c}$~\footnote{In the case for the physical $B_c$ meson, we will
not pursue the further scale separation between $m_c$ and $m_b$,
assuming no significant hierarchy between $x_0$ and $\bar{x}_0$.
Nevertheless in Section~\ref{heavy-light:meson:EM:FF}, we will
explore the consequence of the {\it heavy-quark limit} $m_b\gg m_c$, which
might be relevant for the $B$ meson form factor.}:
\bqa
C^{(0)}_{\rm asym}(Q;m_c,m_b) &=& {4\pi C_F \alpha_s(\mu_R^2) \over Q^2}
\left({e_c \over \bar{x}_0^2}-{e_b \over x_0^2}\right).
\label{NRQCD:short:coeff:LO:physical:Bc:asym}
\eqa
The NRQCD prediction correctly embraces the $1/Q^2$ scaling of
the form factor of a pseudoscalar meson, as it should be.

Without sacrifice of the essential goal,
and for great technical simplicity, it is also useful to imagine a
gedanken experiment by playing with a {\it fictitious} $B_c$ meson,
whose both constitutes carry equal mass: $m_c=m_b$.
In such an idealized case, $x_0=\bar{x}_0={1\over 2}$.
Eq.~(\ref{NRQCD:short:coeff:LO:physical:Bc}) then reduces to~\footnote{Even with $m_b$ set equal to $m_c$,
such a fictitious $B_c$ state, as a flavor non-singlet,
should be distinguished from the flavor-singlet pseudoscalar quarkonium such as $\eta_c$.
The EM transition $\gamma^*\eta_c\to \eta_c$ is simply forbidden due to violation of the $C$-parity.}
\bqa
C^{(0)}(Q; m_c, m_c) &=& 16 \pi C_F e_{B_c^+} \alpha_s(\mu_R^2) {Q^2+ 2 M^2_{B_c}\over Q^4},
\label{NRQCD:short:coeff:LO:fictitious:Bc}
\eqa
where $e_{B_c^+} \equiv e_c-e_b= +1$
is the electric charge of the $B_c^+$ meson.
Its limiting behavior, $C^{(0)}_{\rm asym}(Q; m_c, m_c)=16 \pi C_F \alpha_s/Q^2$,
is particularly simple.

\subsection{NRQCD prediction at NLO in $\alpha_s$}

To assess the NLO perturbative correction to the NRQCD
short-distance coefficient, $C^{(1)}$, one can follow the standard
matching procedure by computing the on-shell quark amplitude
for $\gamma^*+c\bar{b}\to c\bar{b}$ to NLO in $\alpha_s$, neglecting
the relative momentum between $c$ and $\bar{b}$
for both incoming and outgoing $B_c$.
Excluding counter-term diagrams, there are in total 64 NLO Feynman
diagrams. Since several distinct scales: $Q$, $m_b$, $m_c$, will
enter in loop integrals, it is conceivable that fulfilling all the
analytic management would be a highly formidable task.

In this work, we will utilize one of the world-leading
automated \textsc{Feynman Diagram Calculation} packages (FDC),
to fulfill all the required algebras of tensor-reduction and one-loop scalar integral.
FDC is a powerful program based on the symbolic language
\textsc{Reduce}, which was designed to automate the perturbative
quantum field theory calculation in computer. FDC was initially
developed by one of us long ago~\cite{Wang:1993rt}, and the function
of automatic one-loop calculation has recently been realized by Gong and Wang~\cite{Gong:2008:thesis}.
In recent years, the FDC package has been vividly applied to numerous
quarkonium production and decay processes, and has withstood many highly nontrivial tests~\cite{Gong:2007db,Gong:2008ce,Gong:2008sn,Gong:2008ue,Gong:2009kp}.

The masses of the $c$ and $b$ quarks have been retained explicitly
in the calculation. The $n_{lf}=3$ species of
light quark flavors: $u$, $d$, $s$, which occurs in the gluon vacuum
polarization diagrams, have been treated as massless.
Because the masses of the $b$ and $c$ are still much smaller than the momentum transfer $Q$,
one should count the number of {\it active} flavors as $n_{f}=n_{lf}+2=5$.

Both ultraviolet (UV) and infrared (IR) (in our case including soft
and Coulomb) divergences may appear in an individual diagram, while
the would-be collinear divergence has been cutoff by
the heavy quark mass.
Dimensional regularization has been used to regularize both UV
and IR divergences~\footnote{During the calculation we need to
handle the trace of a string of Dirac Gamma matrices containing two
$\gamma_5$ in $D=4-2\epsilon$ dimension. As pointed out in
\cite{Field:1981wx}, one can safely utilize the naive dimensional
regularization prescription to annihilate these two $\gamma_5$.},
and the relative velocity $v$ used to regularize the Coulomb
singularity. All the occurring UV divergences are canceled
systematically by incorporating the counterterm diagrams.
As usual we adopt the $\overline{\rm MS}$ scheme to renormalize the
QCD coupling constant.
One has the freedom to choose some specific renormalization prescriptions for the
quark and gluon fields, {\it e.g.}, $\overline{\rm MS}$ or on-shell scheme.
However, the LSZ reduction formula guarantees that the resulting
short-distance coefficient $C^{(1)}$, which is inferred from the on-shell quark amplitude,
is free from any scheme ambiguity, at least to the NLO which we are considering.

It turns out that the soft IR divergences cancel upon summing all the diagrams, and, as usual,
the remaining Coulomb divergences can be factored into the NRQCD matrix element
through the matching.
Hence the final expression of $C^{(1)}$ becomes both UV and IR finite,
depending only on $Q$, $m_c$, $m_b$, and the renormalization scale, $\mu_R$, respectively.

\begin{figure}[tb]
\begin{center}
\includegraphics*[width=16 cm,angle=0,clip=true]{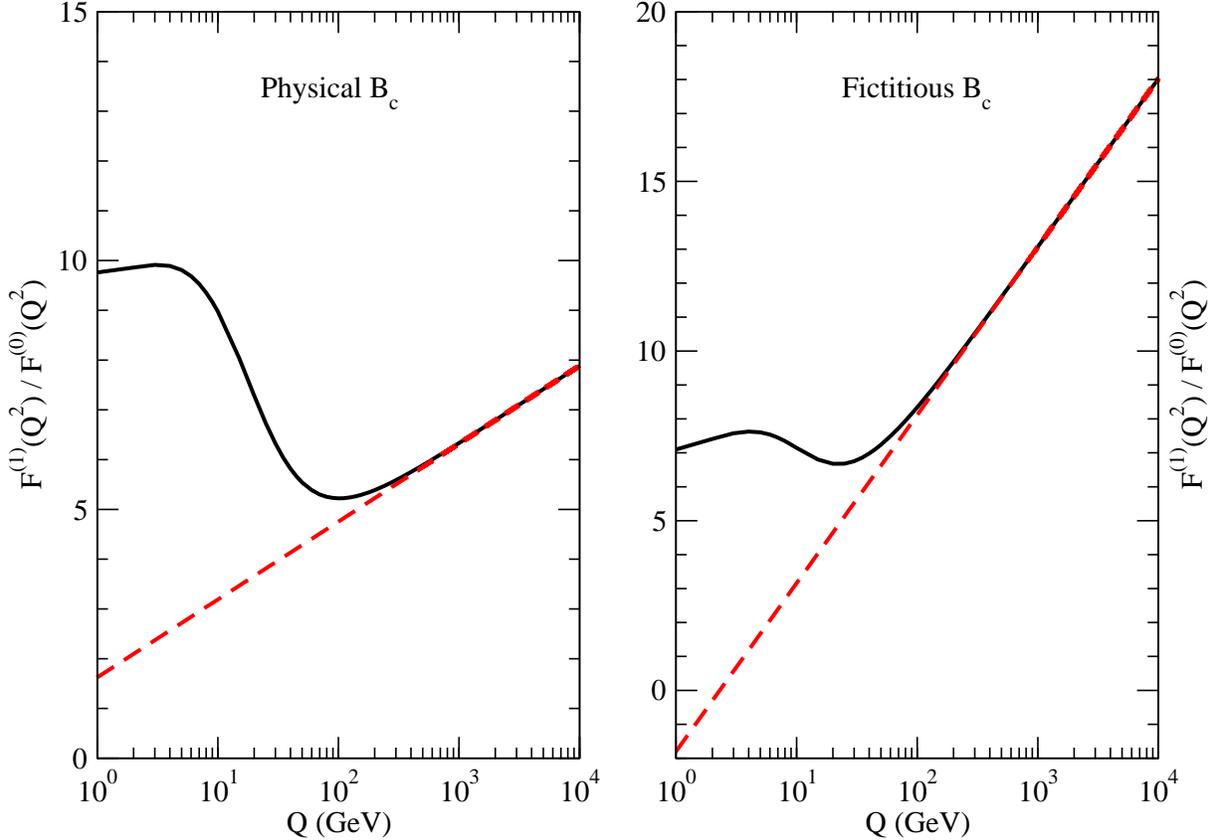}
\caption{
The ratio $F^{(1)}_{B_c}(Q^2)/F^{(0)}_{B_c}(Q^2)$ as a function of $Q$
with $M_{B_c}=6.3$ GeV, $n_f$=5 ($\beta_0={23\over 3})$, and $\mu_R=Q$.
Both the NRQCD and light-cone predictions are shown, where the former is
represented by solid line, and the latter by the dashed line.
The left panel is for the EM form factor of the physical $B_c$ state with
$m_c=1.5$ GeV and $m_b=4.8$ GeV,
while the right panel for a fictitious $B_c$ state with $m_c=m_b=3.15$ GeV.
Numerically, the dashed line in the left panel can be
parameterized by $0.680 \ln Q + 1.624$,
and that in the right panel by $2.152\ln Q - 1.795$, where $Q$ is
in the unit of GeV.
}
\label{Fig:Bc:EM:form:factor}%
\end{center}
\end{figure}

In the case of a physical $B_c$ state with $m_c \neq m_b$, this
calculation poses a rather nontrivial challenge to the capability of
the FDC package, which is perhaps by far the most involved NLO
calculation for the hard exclusive QCD process.
After improving the current algorithm for the scalar-integral part,
and passing several nontrivial internal checks,
FDC can indeed successfully generate the correct result.
Unfortunately, the resulting analytic expression for
$C^{(1)}$ turns to be extremely involved and pathologically lengthy.
There seems absolutely no way to directly manipulate on this
analytic result, not even mentioning to deduce its asymptotic
behavior. Therefore we must be contented with
providing only the numerical results.

In the left panel of Fig.~\ref{Fig:Bc:EM:form:factor}, we explicitly
show the ratio of the NLO form factor to the LO form factor of a
physical $B_c$, $F_{\rm NRQCD}^{(1)}/F_{\rm NRQCD}^{(0)}$ (which
equals to $C^{(1)}/C^{(0)}$) in a wide range of $Q$. For simplicity
we have set $\mu_R=Q$, to eliminate the potentially large UV
logarithm.

The most noteworthy feature is that, after $Q>400$ GeV, the ratio
$F_{\rm NRQCD}^{(1)}/F_{\rm NRQCD}^{(0)}$ starts to grow linearly
with $\ln Q$ (or more precisely, with $\ln(Q/M_{B_c})$ to balance
the dimension). This particular type of single-logarithm scaling is the very
vestige of those would-be collinear divergences that had been cutoff
by the nonzero $b$ and $c$ masses. Within the
confine of NRQCD approach, it is difficult to ascertain the
explicit form of this single logarithm without a complete NLO
calculation. Nevertheless, it has been illustrated that~\cite{Jia:2008ep},
the light-cone approach, can be used effectively to identify, and resum,
such leading collinear logarithms to all orders in $\alpha_s$
for a class of single-quarkonium production processes.
In the following section, we will explicitly see how the
refactorization strategy can help to reproduce this logarithm in a
rather straightforward manner.

In order to acquire a clear view on the asymptotic behavior of
$C^{(1)}$, it is helpful to examine the NLO QCD corrections to the
EM form factor of the aforementioned {\it fictitious} $B_c$ state. The
analytic expression of $C^{(1)}$ in this case, though still involved
to some extent, fortunately becomes far simpler and much more
manageable than that in the case of the physical $B_c$. We note that
the complexity of this calculation is comparable with that of the
NLO correction to $J/\psi+\eta_c$ EM form factor. To our purpose, it
is desirable to deduce the asymptotic behavior of $C^{(1)}$
in a closed form. After some judicious yet tedious manipulations on the
analytic output of FDC, we find the asymptotic expression of
$C^{(1)}$ for this fictitious $B_c$ state to be
\bqa
{C^{(1)}_{\rm asym}(Q;m_c,m_c) \over C^{(0)}_{\rm asym}(Q;m_c,m_c)} &=&
{\beta_0\over 4} \ln {\mu_R^2 \over Q^2} +
{2\over 3}\left(3-2\ln2 \right) \ln{Q^2\over m_c^2}+{4\over 3}\ln^2 2
+{47 \over 18}\ln 2 + {5\over 12}-{2\pi^2\over 9}\,.
\label{NRQCD:C1:C0:fictitious:Bc:asym}
\eqa
Here $\beta_0={11\over 3} N_c-{2\over 3}n_f$ is the one-loop coefficient of the
QCD $\beta$ function. The occurrence of the
$\beta_0\ln\mu_R$ term in (\ref{NRQCD:C1:C0:fictitious:Bc:asym}) is as expected,
required by the $\mu_R$-independence of the NRQCD short-distance coefficients.

In the right panel of Fig.~\ref{Fig:Bc:EM:form:factor}, we have shown
that the ratio $F_{\rm NRQCD}^{(1)}/F_{\rm NRQCD}^{(0)}$  for the fictitious $B_c$ state
over a wide range of $Q$.
The exact NLO prediction is juxtaposed together
with the asymptotic expression, (\ref{NRQCD:C1:C0:fictitious:Bc:asym}).
As indicated clearly in Fig.~\ref{Fig:Bc:EM:form:factor},
as long as $Q> 100$ GeV,
this asymptotic expression starts to converge to the exact NLO result quite well.

Note the coefficient of the collinear logarithm in $C^{(1)}_{\rm asym}$
in (\ref{NRQCD:C1:C0:fictitious:Bc:asym}) is of the algebraic structure $\propto C_F(3-2\ln 2)$,
similar to what appears in the $\eta_c\!-\!\gamma$ transition form factor~\cite{Jia:2008ep,Shifman:1980dk,Sang:2009jc}.
This will be easily understood in the light-cone-based framework in next section.

\section{EM form factor of $\bm{B}_{\bm c}$  in light-cone approach}
\label{Bc:FF:light:cone:fac}

At asymptotically large $Q^2$, both the incoming and outgoing
$B_c$, when viewed in the Breit frame, move nearly with the speed of
light. By virtue of the asymptotic freedom of QCD,
the hard-scattering quark amplitude can be accessed by perturbation theory.
Since both the quark and antiquark inside $B_c$ are dictated by the light-like
kinematics, the hard-scattering amplitude is insensitive to the
small change of the quark mass, $m_{c,b}$, as well as the transverse
momentum carried by the quark and antiquark, $p_\perp$. Thus, the
amplitude can be expanded in powers of $p_\perp$ and $m_{c,b}$,
while the nonperturbative wave function together with the $p_\perp$ and quark-mass-dependent
effects can be lumped into the LCDAs of $B_c$. This picture naturally endorses the
applicability of collinear factorization.

At the leading power in $1/Q$, the EM form factor of $B_c$
can be factored into the convolution of the
perturbatively calculable hard-scattering amplitude $T_H$ with the
leading-twist LCDAs of $B_c$, signified by $\Phi_{B_c}(x)$:
\bqa
F_{\rm LC}(Q^2)&=&\int_0^1 \!\! dx \!\! \int_0^1\!\! dy \,
\Phi^*_{B_c}(y,\mu_F^2)\, T_H(x,y,Q^2,\mu^2_R,\mu^2_F)\,
\Phi_{B_c}(x,\mu^2_F)+ {\mathcal O}(1/Q^4)\,,
\label{leading:twist:factorization:theorem}
\eqa
where $x$, $y$ represent the fractions of light-cone momentum
carried by the $c$ quark in the incident and outgoing $B_c$ states,
and $\mu_R$, $\mu_F$ denote the renormalization and factorization
scales, respectively.

\subsection{Outline of refactorization strategy in light-cone framework}
\label{outline:refactorization}

The factorization theorem (\ref{leading:twist:factorization:theorem}) warrants that,
the hard-scattering kernel, $T_H$, can be systematically improved in perturbation theory.
When computing the hard-scattering amplitude for $\gamma^*+ c(xP)\bar{b}(\bar{x}P)\to c(yP')\bar{b}(\bar{y}P')$
to the leading-power accuracy, the masses and the transverse momenta of the quarks and
antiquarks have been neglected, so both $c$ and $\bar{b}$ move parallel to the momentum of $B_c$.
Obviously, the $T_H$ in our case would be the exactly same as the corresponding one
in the $\pi$ form factor, which has been available for a long while.
It is convenient to organize the hard-scattering amplitude in power series of $\alpha_s$:
\bqa
T_H(x,y,Q^2,\mu_R^2,\mu_F^2)&=&
T_H^{(0)}(x,y,Q^2)+ {\alpha_s(\mu_R^2)\over \pi}
T_H^{(1)}(x,y,Q^2,\mu_R^2,\mu_F^2)+\cdots \,.
\label{TH:decomposition}
\eqa

A crucial ingredient of the factorization formula
(\ref{leading:twist:factorization:theorem})
is the nonperturbative LCDA of the $B_c$ meson.
The $B_c$ LCDA are conventionally parameterized as follows:
\bqa
\Phi_{B_c}(x,\mu_F^2)
 &=& {f_{B_c}\over 2\sqrt{2N_c}}\,\hat{\phi}(x,\mu_F^2),
\label{Bc:LCDA:definition}
\eqa
where $f_{B_c}$ is the (physical) decay constant of $B_c$,
and $\hat{\phi}$ can be viewed as a function characterizing the probability amplitude
for the $c$ ($\bar{b}$) quark to carry the fractional light-cone momentum $x$ ($\bar{x}$).
The factorization scale $\mu_F$ also enters into $\hat{\phi}$,
to preclude those shorter-distance configurations with transverse momentum greater than $\mu_F$
to be erroneously included in the nonperturbative LCDA.
The local limit of the LCDA imposes a model-independent
normalization condition $\int^1_0 \! dx \hat{\phi}(x,\mu_F^2)=1$, valid for any $\mu_F$.

In the case of pion, the distribution amplitude $\hat{\phi}$, which
necessarily probes the long-distance fluctuation of order
$\Lambda_{\rm QCD}$, is a genuinely nonperturbative object.
Therefore it can only be tackled by nonperturbative tools such as
lattice QCD simulation~\cite{Martinelli:1987si,Braun:2006dg}, or by phenomenological methods such as the
QCD sum rules~\cite{Braun:1988qv,Ball:1998je}, or the holographic QCD models~\cite{Brodsky:2006uqa}
that have recently gained much popularity. Unfortunately, despite many intensive
studies for decades, the pion distribution amplitude is still not
accurately known even today.

By contrast, according to the philosophy of refactorization,
the distribution amplitude $\hat{\phi}$ for a quarkonium
may be viewed as the short-distance coefficient (jet) function associated with
matching the quarkonium LCDA onto the NRQCD vaccum-to-quarkonium matrix element,
while the nonperturbative effect of order $mv$ or lower
has been entirely encoded in $f_{B_c}$.
Since heavy quark mass can serve as an infrared cutoff,
this jet function, which encompasses the effects of collinear modes with virtuality of order
$m_{c,b}^2\gg \Lambda_{\rm QCD}^2$,
should be reliably accessible by perturbation theory owing to asymptotic freedom.
At the LO in $v$, this function can be expanded in powers of $\alpha_s$:
\bqa
\hat{\phi}(x,\mu_F^2) &=& \hat{\phi}^{(0)}(x)+ {\alpha_s(\mu_F^2)
\over \pi} \hat{\phi}^{(1)}(x,\mu_F^2)+ \cdots.
\label{Bc:jet:function:expansion}
\eqa
The LO jet function can be trivially inferred,
\bqa
\hat{\phi}^{(0)}(x)=\delta(x-x_0),
\eqa
which simply reflects that both $c$ and $\bar{b}$ partition the
momentum of $B_c$ commensurate to their mass ratios. This is
compatible with the LO NRQCD expansion, in that $c$ and $\bar{b}$
are at rest relative to each other in the $B_c$ rest frame
~\footnote{Note that the LCDA of $B_c$ is asymmetric about
$x={1\over 2}$, which is rather different from the $\pi$ case. The
reason can be easily traced, that is because the flavor symmetry
between heavy quarks $c$ and $b$,  unlike the isospin symmetry between $u$ and $d$,
is badly broken.}.

Separating the nonperturbative LCDA of a meson as the product of the decay constant
and a distribution function $\hat\phi$ in (\ref{Bc:LCDA:definition}),
is merely a parametrization in the case of pion.
Nevertheless, such a parametrization for a quarkonium LCDA,
which is referred to as refactorization in this work,
is a rather useful concept and can lead to quite nontrivial outcome.

In line with the spirit of refactorization,
all the nonperturbative aspects of the twist-2 LCDA of
$B_c$, is evidently contained in the $B_c$ decay constant
$f_{B_c}$. In order to make contact with the NRQCD prediction in
Sec.~\ref{Bc:FF:NRQCD:fac}, it is useful to specify the connection
between the physical $B_c$ decay constant and $f^{(0)}_{B_c}$
introduced in (\ref{decay:constant:LO:alphas}). Beyond the LO in
$\alpha_s$, it is necessary to distinguish $f_{B_c}$ from its
approximate zeroth-order result, $f^{(0)}_{B_c}$. Their connection
is best framed in the NRQCD context, through matching the QCD
axial-vector current to its NRQCD counterpart order by order in
$\alpha_s$ and in $v$. Schematically, the conversion relation
between $f_{B_c}$ and $f^{(0)}_{B_c}$ can be expressed as follows:
\bqa
f_{B_c} &=& f^{(0)}_{B_c} \left( 1+ {\alpha_s(M^2_{B_c})\over \pi} \mathfrak{f}_{B_c}^{(1)} +\cdots \right),
\label{Bc:decay:const:match:NLO:corr}
\eqa
which is valid only at LO in $v$.
For such a static quantity, it is customary and most appropriate to choose the
renormalization scale for $\alpha_s$ appearing in (\ref{Bc:decay:const:match:NLO:corr})
to be around $M_{B_c}$, which characterizes the typical virtuality of quantum fluctuations
integrated out by NRQCD.

Analogous to (\ref{NRQCD:sep:LO:NLO}) in NRQCD factorization,
we also organize the light-cone prediction to
the form factor of $B_c$ in the power series of $\alpha_s$,
\bqa
F_{\rm LC} & = & F_{\rm LC}^{(0)}+
{\alpha_s\over \pi} F_{\rm LC}^{(1)}+\cdots,
\label{Bc:FF:LC:separation}
\eqa

Once the $T_H$, $\hat{\phi}$ and $f_{B_c}$ are separately
known through the NLO accuracy in $\alpha_s$,
following (\ref{leading:twist:factorization:theorem}),
we can then readily recognize the LO and NLO predictions to
the $B_c$ form factor in light-cone perturbation theory.
Obviously, the LO form factor simply assumes the following form:
\bqa
F_{\rm LC}^{(0)} &\sim& \hat{\phi}^{(0)} \otimes T_H^{(0)}\otimes \hat{\phi}^{(0)},
\label{LC:LO:source}
\eqa
where $\otimes$ designates the convolution, and for simplicity we have
suppressed the common multiplicative factors, such as the square of
$f^{(0)}_{B_c}$.

The NLO contribution to the form factor, $F_{\rm LC}^{(1)}$, would receive contributions
from several different sources:
\bqa
F_{\rm LC}^{(1)} &\sim& \hat{\phi}^{(0)} \otimes T_H^{(1)}\otimes\hat{\phi}^{(0)},\;\;
\hat{\phi}^{(1)} \otimes T_H^{(0)}\otimes \hat{\phi}^{(0)},\;\;
\hat{\phi}^{(0)} \otimes T_H^{(0)}\otimes \hat{\phi}^{(1)},\;\;
\mathfrak{f}_{B_c}^{(1)} \,\hat{\phi}^{(0)} \otimes T_H^{(0)}\otimes\hat{\phi}^{(0)}.
\nn
\\
\label{LC:NLO:diff:sources}
\eqa

In (\ref{Bc:FF:LC:separation}), we have deliberately kept some ambiguity on the choice of
the strong coupling constant, that is, we have not specified the scale at which
the $\alpha_s$ should be evaluated.
It is worth noting, however, the different $\alpha_s$ that stem from different NLO sources
in (\ref{Bc:FF:LC:separation}), are in principle affiliated with quite different scales,
{\it e.g.}, $\mu_R$, $\mu_F$, and $M_{B_c}$, as clearly indicated in (\ref{TH:decomposition}),
(\ref{Bc:jet:function:expansion}) and (\ref{Bc:decay:const:match:NLO:corr}).
Hence there seems no rationale to attach a uniform scale to all the occurring $\alpha_s$,
contrary to the form suggested by (\ref{Bc:FF:LC:separation}).

Since the primary goal in this work is to make a critical comparison between
the light-cone and NRQCD approach at the NLO level,
it seems tolerable, at this stage, not to meticulously distinguish the scale attached to each $\alpha_s$.
For the sake of making contact with the NRQCD prediction, and also for simplicity,
from now on we will assume all the $\alpha_s$ in (\ref{Bc:FF:LC:separation}) to be
evaluated at a single scale, say, $\mu_R$.
The error induced by such a mismatch will propagate to the higher order.

At the end of this section (Sec.~\ref{scale:dep:improve}),
we will envisage a more appropriate scale setting scheme,
and briefly address the possible advantage of the refactorization strategy
over the conventional NRQCD factorization calculation
in ameliorating the scale dependence for a fixed-order calculation.

\subsection{Light-cone prediction at LO in $\alpha_s$}

At tree level, there are four Feynman diagrams, which can be obtained by replacing
the shaded ellipse in Fig.~\ref{Fig:sketch:Bc:EM:form:factor}
with a single gluon line between $c$ and $\bar{b}$ quarks,
as well as by attaching the EM current either to the $c$ or the $\bar{b}$ quark line.
Assuming the incident and outgoing $c$ quark to move collinear to the respective $B_c$ mesons,
and taking their corresponding light-cone momentum fractions to be $x$ and $y$, respectively,
it is a straightforward exercise to get the Born-order hard-scattering kernel:
\bqa
T_H^{(0)}(x,y,Q^2)= {16 \pi
C_F \alpha_s(\mu_R^2) \over Q^2} \left( {e_c \over \bar{x} \bar{y}}-
{e_b \over x y}\right).
\label{pion:hard:scatt:part:LO}
\eqa
This expression is known
ever since the pioneering work by Brodsky and Lepage~\cite{Lepage:1980fj}.

As has been pointed out, the LO jet function associated with the $B_c$ LCDA in (\ref{Bc:LCDA:definition})
simply is $\hat{\phi}^{(0)}(x)=\delta(x-x_0)$, compatible with the NRQCD expansion at the zeroth-order in $v$.

Inserting the explicit expressions of $\hat{\phi}^{(0)}(x)$ and the Born-level hard-scattering kernel,
(\ref{pion:hard:scatt:part:LO}), into (\ref{LC:LO:source}),
we obtain the LO prediction to the $B_c$ form factor:
\bqa
F_{\rm LC}^{(0)}(Q^2) &=& {2\pi C_F \alpha_s(\mu_R^2) \over N_c}
{f^{(0)^2}_{B_c} \over Q^2}\left({e_c\over \bar{x}_0^2}-
{e_b \over x_0^2}\right)\,.
\label{Bc:FF:LO:LC}
\eqa
Not surprisingly, one finds the exact agreement between this result and the asymptotic
expression of the LO NRQCD prediction in
(\ref{NRQCD:short:coeff:LO:physical:Bc:asym}).

\subsection{Light-cone prediction at NLO in $\alpha_s$}

One may feel that, the agreement between NRQCD and light-cone predictions at tree level
is easy to envision, and, more or less trivial.
However, concerning the frightening complexity of the NLO calculation in NRQCD side,
one may agree that,
if the equivalence between these two approaches
at NLO can be further established,
it would be an unambiguous indicator that the strategy of
refactorization is on the correct track.

According to the schematic recipe (\ref{LC:NLO:diff:sources}), one can identify
those different components of NLO corrections concretely
and assemble them together.
Consequently, the NLO light-cone prediction to $B_c$ form factor,
$F_{\rm LC}^{(1)}$, can be expressed as
\bqa
F_{\rm LC}^{(1)}(Q^2) &=&
{2\pi C_F \alpha_s(\mu_R^2) \over N_c} {f^{(0)^2}_{B_c} \over Q^2} \Bigg\{
{e_c \over \bar{x}_0^2} \left[
\mathfrak{T}_H^{(1)}
\bigg(x_0,x_0,{\mu_R^2\over Q^2},{\mu_F^2\over Q^2}\bigg)+2 \bar{x}_0 \langle \bar{x}^{-1}\rangle^{(1)}
+2\,\mathfrak{f}_{B_c}^{(1)}\right]
\nn\\
 &-& (
e_c\to e_b, x_0\leftrightarrow \bar{x}_0 ) \Bigg\}\,.
\label{LC:predict:NLO:physical:Bc}
\eqa
$\mathfrak{T}_H^{(1)}$ is related to the NLO hard-scattering
kernel $T_H^{(1)}$, whose precise meaning will be specified in
(\ref{Bc:TH:NLO:hard:kernel}). $\langle \bar{x}^{-1}\rangle^{(1)}$
signifies the first inverse moment of the light-cone
momentum fraction of the $\bar{b}$ quark,
where the superscript implies that $1/\bar{x}$ needs to be folded
with the NLO jet function. Its concrete
expression will be presented in (\ref{inverse:moment:x:xbar:NLO}).

To complete the NLO analysis in light-cone framework, we need to
work out all the involved ingredients in
(\ref{LC:predict:NLO:physical:Bc}) successively.

\subsubsection{NLO correction to the hard-scattering kernel: $T_H^{(1)}$}

One indispensable ingredient for a complete NLO analysis is
the NLO hard-scattering kernel, $T_H^{(1)}$. As
was mentioned, this quantity is exactly identical to the
corresponding one for pion EM form factor,
so there is no need to recalculate it in this work.

The NLO correction to the hard-scattering amplitude for the pion EM
form factor has been investigated by many authors over the last
three decades, but the history of this study seems to have followed
a somewhat twisted path. The NLO calculation was initially carried
out by three independent groups in early 1980s. But unfortunately,
these results did not fully agree with each
other~\cite{Field:1981wx,Dittes:1981aw,Sarmadi:1982yg}. In 1987, after
scrutinizing the previous calculations, Braaten and Tse were able to
locate the origin of the discrepancies among the earlier
works~\cite{Braaten:1987yy}. One decade later, after critically
reexamining all the existing works, Meli\'{c}, Ni\u{z}i\'{c}, and
Passek then presented an ultimately consistent version, and also
conducted a comprehensive phenomenological study by including the
evolution effect of pion LCDA~\cite{Melic:1998qr}. In this work, we
will take the expression of $T_H^{(1)}$ from
Ref.~\cite{Melic:1998qr}.

Prior to quoting the concrete expression, we would like to
recapitulate some noteworthy aspects encountered in this NLO
calculation. Each individual NLO diagram may contain the single UV
pole, as well as the single or double IR pole. The occurring UV
divergences can be eliminated by the standard renormalization
procedure. Conventionally, both the field strength and the strong
coupling constant are renormalized within $\overline{\rm MS}$ scheme.
The single IR pole may be of soft or collinear origin, and the
double IR pole stems from the overlap between soft and collinear
singularities. Upon summing up all the NLO
diagrams, the double IR poles and single soft IR poles cancel, while
only the single collinear IR pole survives in the final expression
of the NLO quark amplitude~\footnote{A comprehensive diagram-by-diagram analysis
for the NLO correction to $\pi$ EM form factor can be found in
Refs.~\cite{Field:1981wx,Melic:1998qr}.}. According to the collinear factorization
theorem, a standard prescription is to employ the $\overline{\rm
MS}$ factorization scheme to absorb this collinear singularity into the pion LCDA,
consequently the ultimate expression for the hard-scattering kernel,
$T_H^{(1)}$, becomes both UV and IR finite. $T_H^{(1)}$ can be
parameterized as
\bqa
T_H^{(1)}(x,y,Q^2,\mu_R^2,\mu_F^2)= {16 \pi
C_F \alpha_s(\mu_R)  \over Q^2} \left\{ {e_c \over \bar{x} \bar{y}}
\mathfrak{T}_H^{(1)}
\bigg(x,y,{\mu_R^2\over Q^2},{\mu_F^2\over Q^2}\bigg)-
{e_b \over x y}  \mathfrak{T}_H^{(1)}
\bigg(\bar{x},\bar{y},{\mu_R^2\over Q^2},{\mu_F^2\over Q^2}\bigg) \right\},
\nn\\
\label{Bc:TH:NLO:hard:kernel}
\eqa
where $\mathfrak{T}_H^{(1)}$ represents the reduced hard-scattering kernel
with EM current attached either to the $c$ or to the $\bar{b}$.
Note $\mathfrak{T}_H^{(1)}$ is a dimensionless quantity.
The explicit dependence of  $\mathfrak{T}_H^{(1)}$ on $\mu_R$ and $\mu_F$
embodies the vestige of those original UV and collinear IR singularities.

To the intended accuracy, it is sufficient for $T_H^{(1)}$ to convolve with two
LO LCDAs $\hat{\phi}^{(0)}$ of $B_c$, which are
$\delta$-functions, therefore suffice it to know the expression of
$\mathfrak{T}_H^{(1)}$ in the special limit $x=y=x_0$. We start from
the analytic expression of the NLO correction to the hard-scattering
kernel tabulated in Ref.~\cite{Melic:1998qr}. After some
straightforward manipulations, and paying particular care to the
spurious singularity affiliated with the limit $x\leftarrow
y$, we end up with a relatively compact expression:
\bqa
& & \mathfrak{T}_H^{(1)}
\bigg(x,x,{\mu_R^2\over Q^2},{\mu_F^2\over Q^2}\bigg) =
{\beta_0 \over 4}\bigg({5\over 3}-2\ln \bar{x}+ \ln{\mu_R^2\over Q^2} \bigg)+
{C_F\over 2} (3+2\ln\bar{x}) \ln{Q^2\over \mu_F^2}+{1\over 3}{\rm Li}_2(\bar{x})
\nn\\
&+& {4\over 3}\ln^2\bar{x}+{1-32 x+157 x^2\over 36 x^2}\ln\bar{x}
+{1\over 6}\ln^2 x +{4-13{\bar x}\over 36 \bar{x}}\ln x+{1-102
x\over 36 x}-{\pi^2\over 36},
\label{Dimensionless:TH:NLO:x:x}
\eqa
which is asymmetric under the interchange $x\leftrightarrow \bar{x}$.
Note this expression is valid specifically for $N_c=3$, and we have not
attempted to keep the full track of the general $N_c$ dependence in all the occurring
color factors. Obviously, the $\ln(Q^2/\mu_F^2)$
term in (\ref{Dimensionless:TH:NLO:x:x}) is the trace of the
collinear IR singularities encountered in the original NLO quark amplitude.

\subsubsection{NLO correction to the jet function: $\hat{\phi}^{(1)}$}

According to (\ref{LC:NLO:diff:sources}), an important class of NLO correction is
from convoluting the tree-level hard-scattering kernel
with one NLO $B_c$ distribution amplitude
and one LO $B_c$ distribution amplitude.
A novel feature in our light-cone treatment is that, the jet function associated with the $B_c$ LCDA
can be systematically improved in perturbation theory.
It is the very feature that renders the strategy of refactorization
practically useful.

Fortunately, the NLO perturbative correction to the jet-function for the
$B_c$ meson, accurate at the LO in $v$ expansion,
is not needed to be computed in this work,
because recently it has already been calculated by Bell and Feldmann~\cite{Bell:2008er}~\footnote{In Ref.~\cite{Bell:2008er},
the authors carried out the calculation by taking $K$ meson as the
prototype for a nonrelativistic system composed by $\bar{s}$ and $u$
quarks, with $m_u\neq m_s$. They performed a separate study for the
$B_c$ meson, assuming that it belongs to the heavy-light meson
family, alike to $B^+$. Since the $B_c$ meson is regarded as a truly
nonrelativistic bound state in the current work, we thus transplant
their expression of $\hat{\phi}^{(1)}$ for the $K$ meson, rather
than theirs for $B_c$.}. Here we just quote their result:
\bqa
& & \hat{\phi}^{(1)}(x,\mu_F^2)=
{C_F\over 2} \left\{\left(\ln{\mu^2_F\over M_{B_c}^2 (x_0-x)^2}-1 \right)
\left[
{x_0+\bar{x}\over x_0-x} {x\over x_0}
\theta(x_0- x) + \left(
\begin{array}{c}
x\leftrightarrow \bar{x}
\\
x_0 \leftrightarrow \bar x_0
\end{array}
\right)\right]\right\}_+
\nn\\
&+& C_F \Bigg\{
\bigg({x \bar{x}\over (x_0-
x)^2}\bigg)_{++} +  {1\over 2}\,\delta^\prime(x-x_0)
\bigg(2 x_0 \bar{x}_0 \ln {x_0 \over \bar{x}_0}+
x_0-\bar{x}_0\bigg)\Bigg\}\,.
\label{Bc:jet:function:NLO}
\eqa
Here the ``+'' and ``++"-prescriptions are understood in the sense
of distributions. For a test function $f(x)$ which has smooth behavior
near $x=x_0$, its convolutions with the ``+'' and
``++"-functions are given by
\begin{subequations}
\bqa
&&\int_0^1 \!\! dx\, [g(x)]_+  f(x)\equiv \int_0^1\!\! dx\,
g(x) \left(f(x)-f(x_0)\right)\,,
\\
&&\int_0^1 \!\! dx\, [g(x)]_{++} f(x) \equiv \int_0^1\!\! dx\,
g(x) \bigg( f(x)-f(x_0)-f^\prime(x_0)(x-x_0)\bigg)\,.
\eqa
\end{subequations}

As one can readily tell from (\ref{Bc:jet:function:NLO}),
the jet function $\hat{\phi}^{(1)}(x)$ has a nonvanishing support in the full range from 0 to 1,
which is in sharp contrast to the infinitely-narrow LO jet function.
In particular, the NLO jet function has developed a long tail,
as a consequence of reshuffling
the momentum fraction between $c$ and $\bar{b}$ to a highly asymmetric configuration
through exchanging energetic collinear gluon. It is interesting
to compare this perturbatively-generated broad profile of the
quarkonium jet function to the phenomenologically determined
quarkonium LCDA, which has a much narrower width of $O(v)$.

As is well known, the evolution of the leading-twist LCDA of a
meson is governed by the renormalization group equation,
which is commonly referred to as Efremov-Radyushkin-Brodsky-Lepage
(ERBL) equation~\cite{Efremov:1979qk,Lepage:1979zb}.
Particularly, the jet function of the $B_c$ meson
obeys the following evolution equation:
\bqa
{d\over d \ln \mu_F^2} \hat{\phi}(x,\mu_F^2)
 &=&
{\alpha_s(\mu_F^2)\over \pi} \, \int^1_0 \! dy \, V_0(x,y)\,\hat{\phi}(y,\mu_F^2)+ O(\alpha_s^2),
\label{BL-evolution:eqn:spin:zero}%
\eqa
where
\bqa
V_0(x,y) &=& {C_F\over
2} \left[{1-x\over 1-y}\left(1+{1\over
x-y}\right)\theta(x-y)+ {x\over y}\left(1+{1\over
y-x}\right)\theta(y-x)\right]_+
\label{BL-kernel-pseudoscalar}%
\eqa
is the corresponding evolution kernel. One can explicitly check
that, upon substituting (\ref{Bc:jet:function:NLO}) into
(\ref{Bc:jet:function:expansion}),
Eq.~(\ref{BL-evolution:eqn:spin:zero}) is indeed satisfied.

With the knowledge of the Born-order hard-scattering amplitude
(\ref{pion:hard:scatt:part:LO}) and the LO expression of the $B_c$
LCDA, ERBL equation can be utilized to identify and resum the
leading collinear logarithms to all order in
$\alpha_s$~\cite{Jia:2008ep}.
In particular, it is not difficult to
apply ERBL equation to infer the single-collinear-logarithm scaling
observed in Fig.~\ref{Fig:Bc:EM:form:factor}.

When convoluting the jet function with the hard-scattering kernel in
the light-cone framework, one often encounters the inverse moment of
$x$ ($\bar{x}$), {i.e.}, the integral of the jet function weighted
by $1/x$ ($1/{\bar x}$), as indicated in (\ref{LC:predict:NLO:physical:Bc}).
With the explicit expression (\ref{Bc:jet:function:NLO}), it is straightforward
to deduce the first inverse moments of $x$ and $\bar{x}$ at NLO accuracy:
\begin{subequations}
\bqa
\langle \bar{x}^{-1}\rangle^{(1)} & \equiv & \int_0^1 \!\! dx\,{\hat{\phi}^{(1)}(x) \over \bar{x}}
= {C_F \over 4 \bar{x}_0} \bigg[(3+2 \ln \bar{x}_0)\ln {\mu_F^2 \over M_{B_c}^2}+
4\,{\rm Li}_2(\bar{x}_0)
\nn\\
&-& 2\ln^2 \bar{x}_0- 2(1+3 \bar{x}_0)\ln \bar{x}_0  -6 x_0\ln x_0 +6 -
{2\pi^2\over 3} \bigg],
\\
\langle
x^{-1}\rangle^{(1)}&\equiv& \int_0^1 \!\! dx\,{\hat{\phi}^{(1)}(x) \over x}=\langle
\bar{x}^{-1}\rangle^{(1)}\big\vert_{x_0\leftrightarrow \bar{x}_0}\,.
\eqa
\label{inverse:moment:x:xbar:NLO}
\end{subequations}
The respective first inverse moments at LO accuracy are trivial, i.e.,
$\langle \bar{x}^{-1}\rangle^{(0)}={1\over \bar{x}_0}$.

The first inverse moment of $\bar{x}$ at NLO depends on
factorization scale logarithmically. It is worth noting that, the
$\ln (\mu_F^2/M_{B_c}^2)$ term has the same coefficient as the
$\ln(Q^2/\mu_F^2)$ term in the NLO hard-scattering kernel in
(\ref{Dimensionless:TH:NLO:x:x}).
This is guaranteed by the general principle of collinear factorization framework.

\subsubsection{NLO correction to $B_c$ decay constant: $\mathfrak{f}_{B_c}^{(1)}$}

As outlined in (\ref{LC:predict:NLO:physical:Bc}),
the last missing piece for a complete NLO analysis is the NLO perturbative
correction to the $B_c$ decay constant. This information can be
inferred through matching the QCD axial vector current onto its
NRQCD counterpart to NLO in $\alpha_s$, which has also been available
long ago~\cite{Braaten:1995ej}:
\bqa
\mathfrak{f}_{B_c}^{(1)} &=& -{3\over 2}C_F + {3\over 4} C_F
(x_0-\bar{x}_0) \ln {x_0\over\bar{x}_0},
\label{Bc:decay:const:matching:coef:NLO}
\eqa
which is symmetric under $x_0\leftrightarrow
\bar{x}_0$.

\subsubsection{Final NLO prediction to $F_{\rm LC}^{(1)}$}

Now it is time to piece all the relevant elements together.
Substituting $\mathfrak{T}_H^{(1)}(x_0,x_0)$ given in
(\ref{Dimensionless:TH:NLO:x:x}), $\langle
\bar{x}^{-1}\rangle^{(1)}$ in (\ref{inverse:moment:x:xbar:NLO}), and
$\mathfrak{f}_{B_c}^{(1)}$ given in
(\ref{Bc:decay:const:matching:coef:NLO}), into equation
(\ref{LC:predict:NLO:physical:Bc}), we finally obtain the complete
NLO light-cone prediction to the $B_c$ EM form factor:
\bqa
& & F_{\rm LC}^{(1)}(Q^2) = {2\pi C_F \alpha_s(\mu_R^2) \over N_c} {f^{(0)^2}_{B_c} \over Q^2}
\Bigg\{
{e_c \over \bar{x}_0^2} \left[
{\beta_0 \over 4}\bigg({5\over 3}-2\ln \bar{x}_0 + \ln{\mu_R^2\over Q^2} \bigg)+
{C_F\over 2} (3+2\ln\bar{x}_0) \ln{Q^2\over M_{B_c}^2} \right.
\nn\\
&+& \left. 3\,{\rm Li}_2(\bar{x}_0)+{1\over 6}\ln^2 x_0 +
{1-32 x_0+37 x_0^2 \over 36 x_0^2} \,\ln \bar{x}_0
+ {4-85 {\bar x}_0 \over 36 \bar{x}_0}\ln x_0 + {1-102 x_0\over 36 x_0}-
{17\,\pi^2\over 36}\right]
\nn\\
 &-& (
e_c\to e_b, x_0\leftrightarrow \bar{x}_0 ) \Bigg\}\,.
\label{Bc:FF:NLO:LC}
\eqa
Note that the factorization scale $\mu_F$ has disappeared from this ultimate
NLO prediction, as it should be. This happens because the IR logarithm
$\ln (Q^2/\mu_F^2)$ in $\mathfrak{T}_H^{(1)}$ smoothly merges with
the UV logarithm $\ln (\mu_F^2/M^2_{B_c})$ in $\langle
\bar{x}^{-1}\rangle^{(1)}$.
The simple manifestation of the collinear logarithms
highlights one attractive power of the light-cone approach.

Equation (\ref{Bc:FF:NLO:LC}) constitutes the climax of this work.
This NLO prediction from the refactorization approach is impressively
compact, in stark contrast with the extremely involved NLO expressions
from the NRQCD approach, though they are supposed to be completely
equivalent at sufficiently large $Q^2$.

\subsection{Comparison of light-cone and NRQCD predictions at NLO in $\alpha_s$}
\label{cmp:lc:NRQCD:NLO}

\subsubsection{Physical $B_c$}

We are now in a position to make a critical comparison between the NRQCD and light-cone
predictions to the $B_c$ EM form factor through NLO in $\alpha_s$.

In the left panel of Fig.~\ref{Fig:Bc:EM:form:factor}, we have depicted the ratio
$F_{\rm LC}^{(1)}/F_{\rm LC}^{(0)}$ as a function of $Q$,
which is generated according to (\ref{Bc:FF:NLO:LC}) and (\ref{Bc:FF:LO:LC}).
The values of $x_0$ and $\bar{x}_0$ are fixed by taking $m_c=1.5$ GeV and $m_b=4.8$ GeV,
relevant for a physical $B_c$ state.
The corresponding NLO NRQCD prediction, has also been juxtaposed in the same plot.
As anticipated, these two predictions do converge together as $Q$ becomes much
greater than $M_{B_c}$.
At $Q=10^2$ GeV, the light-cone asymptote is about 9\% lower than the NRQCD result;
but at $Q=10^4$ GeV, the relative error between the light-cone and NRQCD predictions
shrinks as tiny as $3\times 10^{-5}$.

This highly nontrivial agreement has several important implications.
First, the prediction from the light-cone approach armed with
the refactorization strategy, (\ref{Bc:FF:NLO:LC}),
can be unequivocally regarded as the asymptotic expression of the NLO NRQCD prediction.
This identification is expected to hold order by order in $\alpha_s$.

This agreement also corroborates, in an indisputable manner,
the correctness of the NLO NRQCD calculations presented in this work,
the correctness of the NLO calculation in the light-cone side, {\it i.e.},
the hard-scattering kernel~\cite{Melic:1998qr},
the jet function for the $B_c$ meson~\cite{Bell:2008er},
as well as the NLO correction to the $B_c$ decay constant~\cite{Braaten:1995ej}.

We have stressed that, due to the pathological complexity of the analytic NLO expression
from the NRQCD side, it would be extremely time-consuming, if not impossible,
to deduce the asymptotic behavior directly from this expression itself.
Remarkably, by proceeding along a rather different route,
we are able to deduce, with much ease,
the desired asymptotic NRQCD expression in closed form.

Lying in the heart of the utility of refactorization, is its maximal exploitation of scale
separations. Hard exclusive reaction involving quarkonium is complicated by the coexistence of several
widely-separated energy scales, {\it e.g}., $Q$, $m$, $mv$, and so on.
Refactorization strives to dissect such a multi-scale problem into several simpler steps,
each of which only focuses on a single scale and can thus be tackled easily.
From the example of $B_c$ EM form factor, we hope to have convinced
the readers that this strategy is indeed much more efficient than the
brute-force calculation within the NRQCD factorization.

\subsubsection{Fictitious $B_c$}

To sharpen our understanding, it is also instructive to examine the EM form factor of a
fictitious $B_c$ meson.
We have chosen $m_c=m_b=3.15$ GeV in such a case.
In the right panel of Fig.~\ref{Fig:Bc:EM:form:factor}, the ratio
$F_{\rm LC}^{(1)}/F_{\rm LC}^{(0)}$ for the fictitious $B_c$ state
is also shown for a wide range of $Q$.
As can be clearly seen, the light-cone prediction starts to
overlap with the NRQCD prediction since $Q>10^2$ GeV.
At $Q=10^2$ GeV, the light-cone asymptote is about 3\% lower than the NRQCD result;
but at $Q=10^4$ GeV, the fractional error between these two predictions
reduces to $6\times 10^{-6}$.

For a fictitious $B_c$ meson, the asymptote of the NLO NRQCD predictions has already
been known analytically
in Sec.~\ref{Bc:FF:NRQCD:fac}, thus we can also make a comparison at the analytic level.
Substituting $x_0=\bar{x}_0={1\over 2}$ into (\ref{LC:predict:NLO:physical:Bc}), the
light-cone prediction becomes
\bqa
{F_{\rm LC}^{(1)}(Q^2) \over F_{\rm LC}^{(0)}(Q^2)} &=& \mathfrak{T}_H^{(1)}
\bigg({1\over 2},{1\over 2},{\mu_R^2\over Q^2},{\mu_F^2\over Q^2}\bigg)+\langle
x^{-1}\rangle^{(1)}-4.
\label{LC:predic:ratio:NLO:fictitious:Bc}
\eqa
where the last entity is due to $\mathfrak{f}_{B_c}^{(1)}$ for the equal mass case.
The intended expressions for $\mathfrak{T}_H^{(1)}$ and $\langle
x^{-1}\rangle^{(1)}$ can be simply deduced from
(\ref{Dimensionless:TH:NLO:x:x}) and
(\ref{inverse:moment:x:xbar:NLO})~\footnote{Note that the jet function for the fictitious $B_c$ state
is equal to that for $\eta_c$.
The NLO perturbative correction to the jet function for $\eta_c$ was first evaluated
in \cite{Ma:2006hc}. However, as pointed out in Ref.~\cite{Bell:2008er},
the $\hat{\phi}^{(1)}(x)$ determined in \cite{Ma:2006hc}
does not respect the due normalization condition.}.
The ratio of the NLO prediction to the LO one
for the fictitious $B_c$ state reads:
\bqa
{F_{\rm LC}^{(1)}(Q^2) \over F_{\rm LC}^{(0)}(Q^2)} &=&
{\beta_0\over 4} \left( {5 \over 3}+ 2\ln
2+ \ln {\mu_R^2 \over Q^2}\right)+
{C_F\over 2}\left(3-2\ln2 \right) \ln{Q^2\over M^2_{B_c}}
\nn\\
&& -\,{4\over 3}\ln^2 2 +{25 \over 9}\ln 2- {25\over 9}-{2\pi^2\over 9}\,,
\label{LC:final:NLO:prediction:ratio:fictitious:Bc}
\eqa
where the artificial factorization scale $\mu_F$ cancels.
This expression can also be directly obtained by substituting $x_0={1\over 2}$ in (\ref{Bc:FF:NLO:LC}).
Inserting $\beta_0={23\over 3}$ (for $n_f=5$),
this light-cone prediction, reassuringly, agrees exactly with the
asymptotic NRQCD result tabulated in (\ref{NRQCD:C1:C0:fictitious:Bc:asym}).

From the light-cone perspective, now it should be clear why the coefficient of the collinear logarithm
in the asymptotic NLO NRQCD expressions for the EM form factor of the fictitious $B_c$ and the $\eta_c\!-\!\gamma$ form factor,
appears to be proportional to $C_F \left(3-2\ln2 \right)$. This occurs because
the collinear logarithms in both cases can be identified from
the inverse moment $\langle \bar{x}^{-1}\rangle^{(1)}$.

\subsection{Improving the scale dependence by refactorization strategy}
\label{scale:dep:improve}

In the preceding analysis, our primary goal is to verify that, for the leading-twist hard exclusive
process involving two quarkonia, exemplified by $B_c$ form factor, the light-cone approach,
when armed with the machinery of refactorization,
can be utilized as an efficient and elegant
tool to systematically reproduce the asymptotic NRQCD prediction.
As we have seen, our approach has withstood nontrivial test at the NLO level.
We anticipate that the refactorization method works
presumably to any fixed order in $\alpha_s$.

In higher-order calculation from the NRQCD factorization approach,
it is a common practice to attach all the occurrences of $\alpha_s$ with a single scale,
$\mu_R$ by default.
It has been observed that, for the NRQCD prediction of the double charmonium production
process $e^+e^-\to J/\psi+\eta_c$, the scale dependence does not improve at all
even after including the NLO correction~\cite{Zhang:2005cha,Gong:2007db}.
This may be viewed as a serious drawback of
the conventional NRQCD factorization approach,
presumably attributed to the inadequate disentanglement of
the scales $Q$ and $m$ in NRQCD short-distance coefficients.

In the light-cone approach that implements refactorization,
the complete NLO result encompasses contributions from several different ingredients.
As stressed in Sec.~\ref{outline:refactorization},
the different $\alpha_s$ associated with the different source in (\ref{LC:NLO:diff:sources}),
should in principle be evaluated at different scales,
{\it e.g.}, $\mu_F$, $\mu_R$, and $M_{B_c}$, respectively.
While it might be common to set $\mu_R=\mu_F\sim Q$,
it is certainly more reasonable to evaluate the $\alpha_s$ associated with
the last piece in (\ref{LC:NLO:diff:sources}) at a lower scale around $M_{B_c}$.
By this way, one could, presumably, make a physically more sensible NLO prediction than
the standard NLO NRQCD prediction.

Since our light-cone approach has done a finer job in disentangling the
distinct scales than the standard NRQCD factorization approach,
it is natural to envisage that the scale-dependence might be
significantly reduced with the implementation of the refactorization strategy.
We hope to explicitly illustrate this attractive property
in future publication.

\section{Asymptotic behavior of heavy-light meson EM form factor}
\label{heavy-light:meson:EM:FF}

Thus far, the $B_c$ meson has been treated as a genuine quarkonium state.
However, it has also occasionally been classified
as a heavy-light meson by some authors.
Consequently, we cannot resist the temptation to apply the NLO analysis
performed in previous sections to explore certain features about the
hard exclusive processes involving heavy-light meson.
For definiteness, in below we will take the asymptotic behavior of the
$B^+$ meson form factor as a concrete example.
This section has somewhat digressed from the main thread of this work,
hence an uninterested reader may skip this and go on to next section.

Obviously, NRQCD factorization approach, which is tailor-made to describe
the hard processes involving quarkonium,
will no longer be a rigorous approach to deal with
hard exclusive process involving $B$ meson.
Applying NRQCD factorization approach literally
to analyze the $B$ meson form factor amounts to
modeling the $B$ meson as a nonrelativistic bound state composed of a
heavy $\bar{b}$ quark and a light {\it constitute} $u$ quark~\footnote{For a
phenomenological investigation of time-like $D^{(*)+}D^{(*)-}$ form factor
from this perspective, see \cite{Liu:2003sh}.}.

Provided that $Q^2$ is asymptotically large, the light-cone framework should still apply,
but the strategy of refactorizing the $B$ meson LCDA as the product
of the $B$ meson decay constant and a perturbatively calculable jet function looks
obviously unjustified, since the necessary condition
$m_b, m_u\gg \Lambda_{\rm QCD}$ has been violated.
On the other hand, although modeling the $B$ meson LCDA as $\Phi_{B}(x)
= {f_{B}\over 2\sqrt{2N_c}}\,\delta(x-x_0)$ may seem to be a overly rough approximation,
it may arguably capture some reasonable physics
provided that $m_u$ is taken as the constituent quark mass of order $\Lambda_{\rm QCD}$.
That is, this approximation is compatible with the characteristic picture of a
heavy-light meson,
that the momentum carried by the light quark is predominantly soft.

This said, let us recklessly apply the NRQCD factorization and
our refactorization approach
to the $B$ meson form factor at large $Q^2$,
bearing in mind that a constitute quark model picture has been assumed.
First observe that in the heavy quark
limit  $m_u/m_b \to 0$ (hence $x_0\to 0$ and $\bar{x}_0\to 1$),
the diagrams in which the EM current is attached to the up quark line can be
neglected with respect to those in which the EM current coupled to the $b$ quark,
due to the much higher virtuality of the exchanged gluon in the former case.
Starting from either (\ref{NRQCD:short:coeff:LO:physical:Bc:asym}) or (\ref{Bc:FF:LO:LC}),
one can readily infer the asymptotic expression of the LO $B^+$ form factor:
\bqa
F_{B^+}^{(0)}(Q^2) &=& -{2\pi C_F e_b \over N_c}
{\alpha_s(\mu_R^2) \over Q^2}\left(
{f_B \over x_0}\right)^2\,,
\label{B+:FF:LO:HQET}
\eqa
where $f_B$ is the $B^+$ meson decay constant.
Besides the normal $1/Q^2$ scaling, there is a pronounced enhancement factor
brought in by $1/x_0^2$. Obviously, an extra complication of this case with respect to the $B_c$ form factor,
is that there emerges one additional nonperturbative parameter, $1/x_0$.

At first sight, one may view (\ref{B+:FF:LO:HQET}) as a naive estimate from the nonrelativistic
constitute quark model, which should not be attached with too much significance.

One alternative formalism, yet tailor-made to deal with hard exclusive reaction involving heavy-light meson,
has been developed by Grozin and Neubert some time ago~\cite{Grozin:1996pq}.
They introduced a pair of new $B$ meson nonperturbative distribution amplitudes $\phi_\pm(\omega)$,
specifically defined in the context of heavy quark effective theory (HQET),
where $\omega$ denotes the light-cone energy of the spectator quark.
A nonperturbative parameter that appears in virtually every exclusive $B$
meson decay process, is the first inverse moment of the $B$ meson ``leading-power" distribution amplitude~\footnote{The
term power correction in this context refers to effects suppressed by powers of $1/m_b$,
which should be distinguished with those by powers of $1/Q$.},
the so-called $1/\lambda_B \equiv \int d\omega \phi_+(\omega)/\omega$, which
scales as $1/\Lambda_{\rm QCD}$ by dimensionless counting.

The formalism developed in \cite{Grozin:1996pq} (see also \cite{Beneke:2000wa})
has later found interesting applications in the high-energy $B$ and $D$ meson
production processes~\cite{Braaten:2001bf}.
Analogous to the NRQCD factorization being a proper framework
to describe quarkonium-involved hard reactions,
we may promote the theoretical framework underlying \cite{Grozin:1996pq} as the
{\it HQET factorization} approach,
which may be suitable to tackle the heavy-light-meson-involved hard reactions.

As the NRQCD factorization is based on $v$ expansion,
the appropriate expansion parameter in HQET factorization
is $\Lambda_{\rm QCD}/m_b$.
In \cite{Braaten:2001bf},
a useful connection between the NRQCD model calculation and the
HQET factorization approach was offered:
for a heavy-light meson production process, first
obtaining the reaction amplitude by employing the NRQCD factorization method,
then substituting the singular factor $1/x_0$ by $m_b/\lambda_B$ while neglecting
all the remaining occurrences of $x_0$ elsewhere.
A gratifying fact is that, once such a recipe is adopted in (\ref{B+:FF:LO:HQET}),
one then correctly reproduces the asymptotic LO prediction
made in \cite{Grozin:1996pq}.

\begin{figure}[tb]
\begin{center}
\includegraphics*[width=12 cm, angle=0,clip=true]{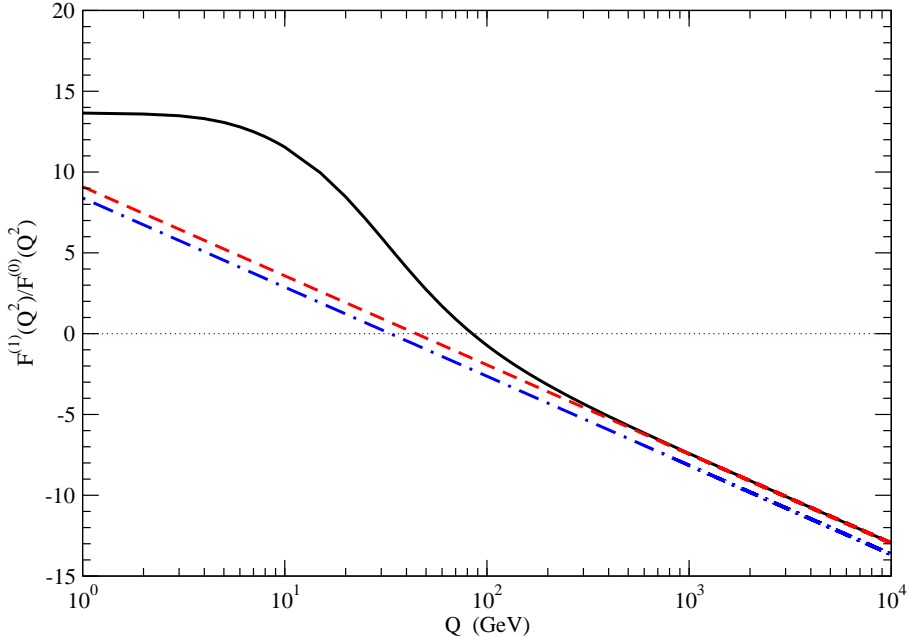}
\caption{
The ratio $F^{(1)}_{B^+}(Q^2)/F^{(0)}_{B^+}(Q^2)$ as a function of $Q$
with $M_{B^+}=5.28$ GeV, $n_f$=5 ($\beta_0={23\over 3})$, and $\mu_R=Q$.
We have chosen $m_u=0.48$ GeV and $m_b=4.8$ GeV.
The NRQCD factorization prediction is represented by the solid line,
while the light-cone predictions from (\ref{Bc:FF:NLO:LC}) by the dashed line.
The dot-dashed line is depicted according to (\ref{Bu:FF:NLO}).
Numerically, the dashed line can be parameterized by $-2.394 \ln Q+9.096$, and the
dot-dashed line by $-2.394 \ln Q+8.393$, where $Q$ is in the unit of GeV.
}
\label{Fig:B+:EM:form:factor}%
\end{center}
\end{figure}

By far, a NLO analysis to the $B$ meson form factor based on the HQET factorization framework
has not emerged yet. Hence it may be interesting to examine the asymptotic behavior of the
$B$ meson form factor at NLO in our NRQCD factorization model and light-cone model.

In Fig.~\ref{Fig:B+:EM:form:factor}, we have shown the NRQCD model prediction and the
corresponding light-cone prediction to the ratio of $F_{B}^{(1)}(Q^2)/F_{B}^{(0)}(Q^2)$
over a wide range of $Q$.
As expected, one observes good agreement between these two approaches after $Q>200$ GeV.
A distinct feature of this ratio is that it decreases as $Q$ increases,
which is opposite to the case of the $B_c$ form factor.

To see such a scaling behavior in a more lucid way,
one may further utilize the heavy quark limit, $m_u/m_b\to 0$,
to simplify the light-cone prediction (\ref{Bc:FF:NLO:LC}).
After some algebra, we find
\bqa
& & {F_{B^+}^{(1)}(Q^2)\over F_{B^+}^{(0)}(Q^2)} =
{\beta_0 \over 4}\bigg({5\over 3}-2\ln x_0 + \ln{\mu_R^2\over Q^2} \bigg)+
{C_F\over 2} (3+2\ln x_0) \ln{Q^2\over M_{B^+}^2}
+ {1 \over 6} \,\ln x_0- {35 \over 12 }-
{17\,\pi^2\over 36}\,.
\nn\\
\label{Bu:FF:NLO}
\eqa
One can see from Fig.~\ref{Fig:B+:EM:form:factor} that such
a limiting behavior is numerically close to the exact NRQCD and
light-cone predictions at large $Q^2$.
From this equation, it can be easily understood
why the collinear logarithm now has developed a negative slope,
because of $m_u/m_b\ll 1$.

Although the prediction (\ref{Bu:FF:NLO}) is far from being rigorous,
it may still contain some essentially relevant ingredient.
For example, the coefficient of the collinear logarithm depends on $\ln x_0$.
Conceivably, this is what one would expect from a NLO calculation in the
HQET factorization, once one identify this term with
the logarithmic moment of the $\phi_+(\omega)$,
the quantity called $\sigma_B$~\cite{Braun:2003wx}.

From our lesson of bridging the NRQCD and
light-cone approaches to effectively describe the quarkonium production,
it may sound appealing to ask whether a similar refactorization strategy can
also be applied to the heavy-light meson production.
That is, is it possible to tie the HQET factorization and the light-cone approaches fruitfully?
Since the $B$ meson LCDA still contains the collinear degrees of freedom of higher virtuality of order $m_b$,
it is natural to expect that such short-distance effect can be separated from the remaining nonperturbative
part. If this reasoning works, the leading-twist $B$ meson LCDA may be, conceivably,
factored into the convolution of the perturbatively calculable jet function
with the $\phi_+(\omega)$.
One attractive point is that, such a refactorization program enables one to
manifestly disentangle the perturbative collinear logarithm of $\ln(Q/M_B)$ from the
nonperturbative soft logarithm of $\ln(M_B/\Lambda_{\rm QCD})$.
While the former logarithm is controlled by the ERBL equation,
one can invoke Lange-Neubert evolution equation~\cite{Lange:2003ff}
to deal with the latter.

\section{Challenge of the refactorization program to $\bm{J}\bm{/}\bm{\psi}\bm{-}\bm{\eta}_{\bm c}$ EM form factor}
\label{chall:Jpsi+etac:EM:FF}

In recent years, one of the most widely-studied
hard exclusive reactions is perhaps the double charmonium production process
$e^+e^-\to J/\psi+\eta_c$,
which was observed at the $B$ factories a number of years ago~\cite{Abe:2002rb,Aubert:2005tj}.
This process provides a powerful probe to extract the  $J/\psi+\eta_c$
EM form factor in the time-like region.
One can define this form factor, which be referred to as $G(Q^2)$ hereafter,
as follows:
\bqa
\langle J/\psi(P,\epsilon(\lambda))+\eta_c(P^\prime)\vert
J_{\rm em}^\mu \vert 0 \rangle = i\,G (Q^2)\,\epsilon^{\mu\nu\rho\sigma}
P_\nu P^\prime_\rho \epsilon^*_\sigma(\lambda)\,,
\label{EMFM:jpsi+etac}
\eqa
where $Q=P+P^\prime$, and $\epsilon(\lambda)$ represents the polarization vector for the $J/\psi$
with helicity $\lambda$. This specific Lorentz structure is constrained by the Lorentz and parity
invariance, from which one can easily see
that the outgoing $J/\psi$ must be transversely polarized, i.e., $\lambda=\pm 1$.

In analogy with (\ref{LO:NRQCD:factorization:formula}), according to the NRQCD factorization,
one can express the form factor $G(Q^2)$ as
\bqa
G_{\rm NRQCD}(Q^2)&=& C(Q;m_c)  {\langle J/\psi(\lambda)| \psi^\dagger {\bm \sigma}\cdot
{\bm\epsilon^*}(\lambda)\,\chi| 0 \rangle  \over
\sqrt{2 N_c m_c}} \cdot {\langle \eta_c | \psi^\dagger \chi| 0 \rangle\over \sqrt{2N_c m_c}}
+ {\mathcal O}(v^2)\,,
\label{LO:v:NRQCD:fact:formula:Jpsi+etac}
\eqa
where we have only retained the contribution at LO in $v$.

The NRQCD short-distance coefficient, $C(Q;m_c)$,
can again be organized in power series of the strong coupling constant,
$C=C^{(0)}+ {\alpha_s \over \pi} C^{(1)}+\cdots$.
There are four lowest-order Feynman diagrams for hard-scattering process,
and the tree-level result of $C^{(0)}$ is~\cite{Braaten:2002fi}
\bqa
C^{(0)}(Q;m_c) &=&  256 \pi  e_c  C_F \alpha_s(\mu_R^2) \, {m_c\over (Q^2)^2}.
\label{NRQCD:short:coeff:LO:Jpsi+etac}
\eqa
As is well known, since this process necessarily violates the helicity selection rule~\cite{Brodsky:1981kj},
so the form factor has to be suppressed by extra factors of $1/Q$ relative to the $B_c$ form factor.
From (\ref{NRQCD:short:coeff:LO:Jpsi+etac}) one can infer that the helicity conservation
is violated by the heavy quark mass.

\begin{figure}[tb]
\includegraphics*[width=14 cm,angle=0,clip=true]{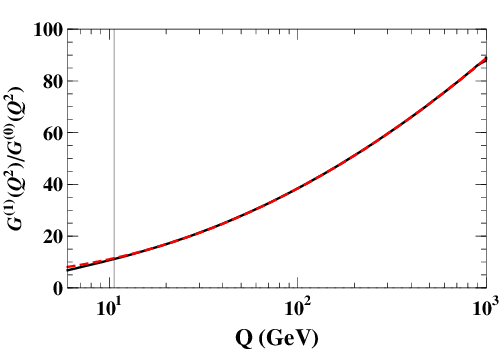}
\caption{The ratio of time-like $J/\psi+\eta_c$ EM form factor
$G^{(1)}(Q^2)/G^{(0)}(Q^2)$ as a function of $Q$. We have adopted
$m_c=1.5$ GeV, $n_f$=4 ($\beta_0= {25\over 3}$), and $\mu_R=Q$. Both
of the exact and asymptotic NLO results in NRQCD factorization
framework are shown, where the former is represented by the solid
line, and the latter [defined in (\ref{NRQCD:C1:C0:Jpsi:etac:asym})]
by the dashed line. The vertical line marks the place at $Q\equiv
\sqrt{s}=10.58$ GeV of $B$ factory energy. Numerically, the dashed
line can be parameterized by $2.167 \ln^2 Q - 3.077 \ln Q + 6.672$,
where $Q$ is in the unit of GeV.}
\label{Fig:Jpsi+etac:EM:form:factor}%
\end{figure}

The NLO perturbative correction to this process was first computed in NRQCD factorization
context in \cite{Zhang:2005cha}, in which only the numerical result at
$Q \equiv \sqrt{s}=10.58$ GeV of $B$-factory energy was presented.
This calculation was later redone by Gong and Wang, as the maiden application of
the FDC package~\cite{Gong:2007db}.
They confirmed the result given in \cite{Zhang:2005cha},
in addition, they also provided the fully analytical expression for $C^{(1)}$.
This information is certainly useful, since one can then acquire a global understanding
of this form factor at asymptotically large $Q^2$.
After some judicious but tedious manipulations on the analytic expressions assembled
in \cite{Gong:2007db},
we can extract the asymptotic behavior of this form factor at NLO in $\alpha_s$:
\bqa
{{\rm Re}[C^{(1)}_{\rm asym}(Q)] \over C^{(0)}_{\rm asym}(Q)} &=&
{13\over 24} \ln^2{Q^2\over m_c^2}-
{41\over 24} (2\ln 2-1)\ln {Q^2\over m_c^2}+ {\beta_0\over 4} \ln {\mu_R^2\over Q^2}
\nn\\
&+& {71\over 8}\ln2+ {59\over 24}\ln^2 2 - {23\over 18}- {\pi^2\over
36}.
\label{NRQCD:C1:C0:Jpsi:etac:asym}
\eqa
where we have normalized $C^{(1)}$ with respect to $C^{(0)}$, and
suppressed the imaginary part for simplicity. In
Fig.~\ref{Fig:Jpsi+etac:EM:form:factor}, we have depicted the ratio
of the time-like $J/\psi+\eta_c$ EM form factors
$G^{(1)}(Q^2)/G^{(0)}(Q^2)$, or equivalently,
$C^{(1)}(Q^2)/C^{(0)}(Q^2)$, as a function of $Q$, including both
the exact and asymptotic NLO results. It is interesting to observe
that, unlike in the $B_c$ case, the asymptotic expression starts to
decently reproduce the exact NLO result at rather low scale of $Q$,
that is, the mutual agreement at $B$-factory energy seems already
quite satisfactory~\footnote{To be concrete, at $Q= 10.58$ GeV, one
finds that ${\rm Re} C^{(1)}_{\rm asym}/C^{(0)}_{\rm asym} = 11.47$,
which is only 3\% larger than the exact NLO result $11.15$.}.

A peculiar feature is that, as manifested in Fig.~\ref{Fig:Jpsi+etac:EM:form:factor},
here the leading scaling behavior of the NLO correction
is actually governed by the {\it double} logarithm of type $\ln^2(Q^2/m_c^2)$~\footnote{It is
interesting to note that ${\rm Im}[C^{(1)}_{\rm asym}(Q)]/ C^{(0)}_{\rm asym}(Q)$, whose full analytic
expression is not given here, contains the single logarithm $-\pi{13\over 12} \ln {Q^2\over m_c^2}$.
Conceivably, with this input, one can reconstruct this double logarithm in the real part in
(\ref{NRQCD:C1:C0:Jpsi:etac:asym}) via the dispersion relation.}.
This is in stark contrast to the asymptotic behavior of
the $B_c$ form factor, as given in (\ref{Bc:FF:NLO:LC}) and
(\ref{LC:final:NLO:prediction:ratio:fictitious:Bc}), whose leading behavior at NLO in $\alpha_s$
is represented by the {\it single} collinear logarithm only.

It is interesting to investigate the numerical significance of this
double logarithm term. Setting $\mu_R=Q$ and $m_c=1.5$ GeV in
(\ref{NRQCD:C1:C0:Jpsi:etac:asym}), we find the value of the double
logarithm ${13\over 24} \ln^2{Q^2\over m_c^2}$ equals $8.27$ at $Q=
10.58$ GeV, which already constitutes 72\% of the full asymptotic
result $ {\rm Re} C^{(1)}_{\rm asym}/ C^{(0)}_{\rm asym}= 11.47$.
The dominance of this double logarithm becomes very prominent at
higher energy. For example, if the $e^+e^-$ center-of-mass energy is
chosen to be near the $Z^0$ pole, i.e., $Q=91.19$ GeV, one finds
that the double logarithm term becomes $36.55$, which is 99\% of the
the full asymptotic result $ {\rm Re} C^{(1)}_{\rm asym}/
C^{(0)}_{\rm asym}= 36.91$. This numerical study undoubtedly
indicates that, the importance the double logarithm term may
severely ruin the stability of the fixed-order perturbative
expansion, and in order to obtain the controlled prediction, it
seems compulsory to resum these types of logarithms in NRQCD
short-distance coefficient to all orders in $\alpha_s$, which is
perhaps relevant even at $B$ factory energy.

At first sight, the occurrence of this double logarithm seems to be
at odds with the general principle of light-cone framework,
especially our refactorization program. Recall that when computing
the $O(\alpha_s)$ quark amplitude for $\pi$ form factor, the double
IR poles and the single IR soft poles have canceled away upon
summing all the NLO diagrams, and only the single collinear (mass)
singularities can survive in the final answer.

In passing, we note that the rising of double logarithm are not something completely new
for the double quarkonium production processes in NRQCD factorization approach.
In particular, this type of scaling behavior has already been encountered in the
exclusive processes such as $\Upsilon\to J/\psi+\eta_c$~\cite{Jia:2007hy}
and $\eta_b\to J/\psi+J/\psi$~\cite{Gong:2008ue,Sun:2010qx}.

One can quickly see that, all the aforementioned double-quarkonium reactions
share a common trait, i.e., that the celebrated helicity selection rule has
been violated in all of them.
This implies that the asymptotic scaling behaviors of
these reactions are suppressed by powers of $1/Q$ with
respect to that of the $B_c$ form factor. Thus to make a leading
nonvanishing prediction for such type of processes in light-cone
formalism, one necessarily needs to include
the {\it higher-twist} LCDAs of quarkonium.

Unfortunately, a long-standing difficulty associated
with these higher-twist LCDAs in collinear factorization framework is that,
one will encounter the ubiquitous {\it end-point singularity} when
convoluting these LCDAs with the hard kernel~\cite{Chernyak:1983ej}.
By far there is no
universally accepted recipe to remedy this notorious problem other
than some ad hoc phenomenological parametrization~\footnote{The so-called zero-bin subtraction
method~\cite{Manohar:2006nz} has
been recently proposed to solve the endpoint singularity problem.
But there seems to exist some controversy about its validity.}.
For example, there seems no consistent way to investigate the NLO correction for the $\rho\pi$ EM form factor
in the light-cone formalism.
By the same token, this may cast some shadows on the solidity of those phenomenological studies on
$\gamma^*\to J/\psi+\eta_c$ from the light-cone approach~\cite{Ma:2004qf}.

We are thus facing a dilemma to carry out the refactorization procedure to the processes like
$\gamma^*\to J/\psi+\eta_c$, because of our incapability of making a consistent NLO analysis of this
higher-twist reaction in the light-cone framework~\footnote{Recently there has come out an
all-order-in-$\alpha_s$ factorization proof for the processes of
$e^+e^-$ annihilation into double quarkonia~\cite{Bodwin:2008nf},
which seems to largely build upon the arsenal of collinear factorization.
The key technique in their proof is the soft and collinear approximations of QCD interaction which conserve the quark
helicity. Hence their proof may be regarded as applicable to those leading-twist (helicity-conserving) reactions
such as $B_c$ EM form factor.
For higher-twist (helicity-suppressed) processes such as
$e^+e^-\to J/\psi+\eta_c$, one has to include the spin-dependent interactions to flip the quark helicity,
which is necessarily beyond the soft and collinear approximations employed
in \cite{Bodwin:2008nf}.
Therefore it is not clear to us
whether their proof can be safely applied to
$e^+e^-\to J/\psi+\eta_c$ or not.}.

\begin{figure}
\begin{center}
\includegraphics[height=11 cm]{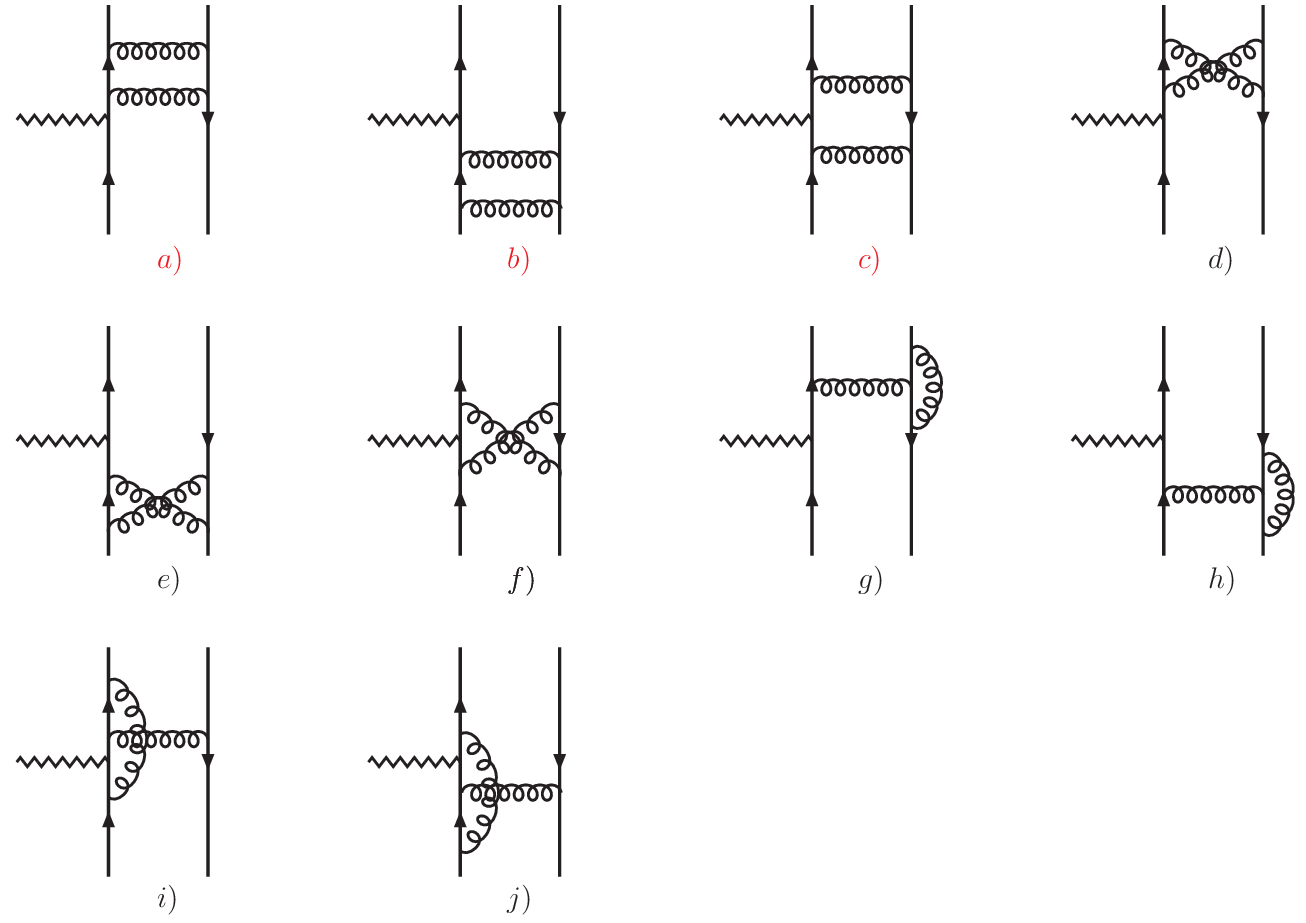}
\caption{
The NLO diagrams that contain the double logarithm $\ln^2(Q^2/m_c^2)$ (in Feynman gauge)
for the process $\gamma^*\to J/\psi+\eta_c$.
The respective charge-conjugated diagrams have been omitted.
}
\label{Fig:diagram:jpsi:etac:double:log}
\end{center}
\end{figure}

Notwithstanding the theoretical limitation of the light-cone
approach, there is still something worth learning solely based upon
the NLO NRQCD results. After a diagram-by-diagram anatomy of the FDC
output~\cite{Gong:2007db}, one is able to single out those diagrams
that contain double logarithms, which are depicted in
Fig.~\ref{Fig:diagram:jpsi:etac:double:log}. It is easy to observe
that the subset of ladder diagrams
[Fig.~\ref{Fig:diagram:jpsi:etac:double:log}$a)$ through $c)$] has a
color structure $\propto  {\rm tr}(T^a T^a T^b T^b)\propto C_F^2$,
while the remaining subset of diagrams has a different color
structure $\propto{\rm tr}(T^a T^b T^a T^b)\propto (C_F-{1\over
2}C_A)C_F$, where $C_A=N_c$ is the Casmir for the adjoint
representation of $SU(N_c)$. One can examine that the sum of all the
diagrams in each subset still contains a nonvanishing double
logarithm. As a result, for an arbitrary $N_c$, the coefficient of
the double logarithm appearing in (\ref{NRQCD:C1:C0:Jpsi:etac:asym})
is a linear combination between $C_F^2$ and $C_F C_A$. This
color-factor dependence differs from the standard Sudakov double
logarithm.

It is instructive to reexamine the same topologies of diagrams
in Fig.~\ref{Fig:diagram:jpsi:etac:double:log} for the case of $\pi$ form factor.
As is illustrated in \cite{Field:1981wx,Melic:1998qr}, in that case
the diagrams Fig.~\ref{Fig:diagram:jpsi:etac:double:log}$a)$ through $c)$ only contain a {\it single} IR pole,
in contrast to the case of $\gamma^*\to J/\psi+\eta_c$.
Furthermore, for the remaining diagrams in Fig.~\ref{Fig:diagram:jpsi:etac:double:log},
even though each of them still contains double IR pole,
their sum does not.
This comparative study clearly shows that
the double logarithms are sensitive to the helicity structure of the specific process.

It is perhaps more transparent to trace the origin of these double logarithms
by employing the {\it method of region}~\cite{Beneke:1997zp}.
Following the version of refactorization outlined in Ref.~\cite{Jia:2008ep},
one may identify the relevant degrees of freedom in the
NRQCD short-distance coefficient $C^{(1)}$. There is always a ``\emph{hard}" region,
in which the loop momentum scales as $p^\mu\sim Q$. In this region,
the mass of heavy quark should be treated
as a small perturbation. There are several kinds of ``infrared" modes,
\emph{soft} ($p^\mu\sim m$), \emph{collinear} ($p^+\sim Q, p^-\sim
m^2/Q, p_\perp\sim m$), and \emph{anti-collinear} ($p^+\sim m^2/Q,
p^-\sim Q, p_\perp\sim m$).
The mass of the $c$ quark must be
retained in these lower-energy regions. The validity of NRQCD
factorization guarantees there is no overlap between these
``infrared" quanta and those truly infrared modes intrinsic to NRQCD
such as the potential mode. In the case of $B_c$ form factor,
the leading contributions only arise from the hard and collinear
(or anti-collinear, but not simultaneously both) regions, while
the soft region cannot make a net contribution at leading power.
In the $J/\psi+\eta_c$ case,
it is expected that each of these four regions will make a contribution
in the leading nonvanishing power (of course, there should be an overall power-suppressed factor).
The double logarithm should originate from the overlap between the (anti-)collinear and soft regions.
In this sense, we may dub this type of double logarithm as the
{\it power-suppressed Sudakov logarithm}.

As has been expounded in \cite{Beneke:2003pa}, wherever the double
logarithm appears, the respective loop integrals in the two
overlapping regions become ill-behaved separately. Rather one needs
to introduce the extra regulator besides the dimensional
regularization for each individual region, such as the {\it
analytic} regularization. This artificial regulator will be
eliminated only after summing the contributions of both regions.
This symptom signals the breakdown of the naive collinear-soft
factorization, which is the very cause for the endpoint singularity.
In phenomenological study of the $\rho\pi$(or $J/\psi\eta_c$) EM
form factor in light-cone approach at LO, one may deliberately
choose the profile of the LCDAs such that it falls off sufficiently
fast near the endpoint, as an means to circumvent the end-point
singularity problem (for example, see
\cite{Braguta:2008hs,Ma:2004qf}). However, the real problem is that,
this symptom is deeply rooted in the breakdown of the collinear
factorization for higher-twist reaction, which cannot be simply
overcome by invoking some phenomenological trick.

Even though it is feasible to reproduce the asymptotic behavior in
(\ref{NRQCD:C1:C0:Jpsi:etac:asym}) with the aid of the method of
region, one perhaps still can not proceed far without a general
guidance of a valid factorization theorem. For instance, it still
remains to be a harsh challenge to identify and resum these
power-suppressed Sudakov logarithms to all orders in $\alpha_s$.
Perhaps a satisfactory answer to this question demands that one
first resolves the end-point singularity problem in a consistent
manner. This direction definitely deserves further exploration.

\section{Summary}
\label{summary}

It is an indisputable fact that both the NRQCD factorization and the
collinear factorization approaches have their
own strengths and limitations in describing hard exclusive reactions involving heavy quarkonium.
In this work, we have illustrated how to coherently tie these two approaches
to achieve the optimized predictive power.
This is made possible by invoking the strategy of {\it refactorization}, e.g.,
by further factoring the quarkonium LCDA into a sum of the products of the
universal yet perturbatively-calculable jet functions and the nonperturbative
vaucuum-to-quarkonium NRQCD matrix elements
(i.e., quarkonium decay constant).
Through a comprehensive comparative study, we have verified that,
for a class of hard exclusive process involving two quarkonium,
exemplified by the $B_c$ form factor at large $Q^2$,
the light-cone approach with refactorization can be utilized to reproduce
the (leading-twist) asymptotic
result of NRQCD factorization prediction, at the NLO in $\alpha_s$ while at the LO in $v$.
We hope to have convinced the readers that
this refactorization program is much simpler,
and, more efficient,
than the brute-force NRQCD-factorization-based calculation.

Quite conceivably, for most realistic quarkonium production processes, {\it e.g.},
$\eta_c\!-\!\gamma$ transition form factor to be the simplest,
it is beyond our current technical capability to analytically
investigate the next-to-next-to-leading (NNLO) perturbative correction
in NRQCD factorization formalism.
Fortunately, the refactorization approach,
owing to its maximally disentanglement of the scales,
will serve as the indispensable, and,
perhaps the only viable, calculational device to fulfill this goal.
Beside such a remarkable technical advantage,
the strategy of refactorization also helps one to gain better theoretical control,
{\it e.g.}, to resum collinear logarithms to all orders in $\alpha_s$, and
to significantly reduce the scale dependence for a fixed order calculation.
We hope to explicitly show these attractive features in future publication.

There are quite a few leading-twist quarkonium production processes
for which the refactorization program may be of interesting applications, {\it e.g.},
$B$ meson exclusive decay to a $S$-wave charmonium plus a light meson,
and $\gamma\gamma\to B_c^+ + B_c^-$.
It is also worth further factorizing the quarkonium LCDA to higher order in $v$ expansion,
by which one is then capable of elegantly reproducing the NRQCD predictions for
$P$-wave quarkonium production or relativistic correction
to $S$-wave quarkonium production.

Another interesting direction is to pursue a similar refactorization program
in hard exclusive reactions involving a heavy-light meson,
e.g., to further factorize the leading-twist $B$ meson LCDA into the convolution of
a perturbatively-calculable yet universal jet function
with the $B$ meson distribution amplitude specifically defined
within the HQET context.

It is an empirical fact that our refactorization program confronts
serious obstacle when applied to {\it helicity-suppressed}
double-quarkonium production processes,
such as $\gamma^*\to J/\psi+\eta_c$, $\eta_b\to J/\psi+J/\psi$, and $\Upsilon\to J/\psi+\eta_c$.
We believe that the double logarithm appearing in the NLO NRQCD
short-distance coefficients in these processes is intimately linked with the
end-point singularity problem in light-cone approach.
In our opinion, a satisfactory control of this double-logarithm for these processes
will offer a pivotal insight into the ultimate solution to this
long-standing problem in collinear factorization.

\begin{acknowledgments}
One of the authors, Y.~J., wishes to thank Kexin Cao, for her kind
assistance in deducing the asymptotic expression of the NLO NRQCD
results for the $J/\psi+\eta_c$ EM form factor, and more
importantly, for the wonderful gift sent to him during the long
period of preparing this manuscript: a lovely son.
D.~Y. thanks Martin Beneke for discussions on the calculation of quarkonium LCDA.
We are also indebted to Xiu-Ting Yang for anatomizing the FDC output
for the $e^+e^-\to J/\psi+\eta_c$ process, and sorting out those NLO
diagrams that contain the double logarithms.
We thank Geoff Bodwin for discussions on the nature of these double
logarithms.
We are also grateful to Hai-Rong Dong and Feng Feng for pointing out
a minor error in both exact and asymptotic NLO expressions for the
$J/\psi+\eta_c$ EM form factor in the earlier version of the paper
(i.e., the ${\mathcal O}(1)$ constant $-{17\over 18}$ in equation
(\ref{NRQCD:C1:C0:Jpsi:etac:asym}) of previous version should be
replaced with $-{23\over 18}$), and for regenerating
Fig.~\ref{Fig:Jpsi+etac:EM:form:factor} for us.
This research was supported in part by the National Natural Science
Foundation of China under Grants No.~10875130, No.~10979056,
No.~10705050 and No.~10935012.
\end{acknowledgments}


\end{document}